\documentclass[hackGuillaume]{svjour3} % hackGuillaume natbib
\usepackage[utf8]{inputenc}
\usepackage[T1]{fontenc}
\usepackage[english]{babel}
\usepackage{times,mathptmx}
\usepackage{latexsym} % $\rhd$
\usepackage{amsmath,amssymb}
\usepackage{nicefrac}
\usepackage{relsize}

\usepackage{multicol}
\usepackage{siunitx}
\usepackage{float}
\usepackage[caption=false]{subfig}
\usepackage{booktabs}

\usepackage{graphicx}
\usepackage[usenames,svgnames,table]{xcolor}
\usepackage[colorlinks,allcolors=MediumBlue,breaklinks,hyperfootnotes=true,
	pdftitle={Tortured phrases: A dubious writing style emerging in science},
	pdfauthor={Guillaume Cabanac, Cyril Labbé, Alexander Magazinov},
	pdfsubject={Preprint},
	pdfkeywords={AI-generated texts, GPT, Misconduct, Research integrity, Tortured phrases}
]{hyperref}

\usepackage{tcolorbox}

\newcommand{\supplmat}{See \hyperlink{apx:supplmat}{Appendix}}

\newcommand{\fakeframcol}{red!20!white}
\newcommand{\fakebackcol}{red!5}
\newcommand{\trueframcol}{green!20!white}
\newcommand{\truebackcol}{green!5}

\newcommand{\faketbox}[2]{
{
\small
\begin{tcolorbox}[colback=\fakebackcol,colframe=\fakeframcol,coltitle=black,title={#1}]
#2%
\end{tcolorbox}
}
}

\newcommand{\truetbox}[2]{
{
\small
\begin{tcolorbox}[colback=\truebackcol,colframe=\trueframcol,coltitle=black,title={#1}]
#2%
\end{tcolorbox}
}
}

\usepackage[natbibapa]{apacite}    % instead of passing natbib to svjour3
\newcommand{\doi}[1]{\href{https://doi.org/#1}{#1}}
\usepackage{underscore}
\begin{document}
\journalname{arXiv Preprint}

\title{Tortured phrases: A dubious writing style emerging in science}
\subtitle{Evidence of critical issues affecting established journals}

\author{Guillaume Cabanac \and Cyril Labbé \and Alexander Magazinov}
\authorrunning{G. Cabanac et al.}

\institute{
All parts of this work are contributed jointly by all authors. Authors are listed in alphabetical order.\\[8pt]
G. Cabanac\at
University of Toulouse, Computer Science Department, IRIT UMR 5505 CNRS, 31062 Toulouse, France\\
\email{guillaume.cabanac@univ-tlse3.fr}\\
ORCID: \href{http://orcid.org/0000-0003-3060-6241}{0000-0003-3060-6241}
\and
C. Labbé\at
Univ. Grenoble Alpes, CNRS, Grenoble INP, LIG, 38000 Grenoble, France\\
\email{cyril.labbe@univ-grenoble-alpes.fr}\\
ORCID: \href{http://orcid.org/0000-0003-4855-7038}{0000-0003-4855-7038}
\and
A. Magazinov\at
Yandex, 82 Sadovnicheskaya str., Moscow 115035, Russia\\
\email{magazinov-al@yandex.ru}\\
ORCID: \href{https://orcid.org/0000-0002-9406-013X}{0000-0002-9406-013X}
}

% Started writing on 30-APR-2021
\date{Version: July 12, 2021}
\maketitle 

\urlstyle{same}

\begin{abstract}
    Probabilistic text generators have been used to produce fake scientific papers for more than a decade.
    Such nonsensical papers are easily detected by both human and machine.
    Now more complex AI-powered generation techniques produce texts indistinguishable from that of humans and the generation of scientific texts from a few keywords has been documented.
    Our study introduces the concept of \emph{tortured phrases}: unexpected weird phrases in lieu of established ones, such as `counterfeit consciousness' instead of `artificial intelligence.'
    We combed the literature for tortured phrases and study one reputable journal where these concentrated en masse.
    Hypothesising the use of advanced language models we ran a detector on the abstracts of recent articles of this journal and on several control sets.
    The pairwise comparisons reveal a concentration of abstracts flagged as `synthetic' in the journal.
    We also highlight irregularities in its operation, such as abrupt changes in editorial timelines.
    We substantiate our call for investigation by analysing several individual dubious articles, stressing questionable features: tortured writing style, citation of non-existent literature, and unacknowledged image reuse.
    Surprisingly, some websites offer to rewrite texts for free, generating gobbledegook full of tortured phrases.
    We believe some authors used rewritten texts to pad their manuscripts.
    We wish to raise the awareness on publications containing such questionable AI-generated or rewritten texts that passed (poor) peer review.
    Deception with synthetic texts threatens the integrity of the scientific literature.
\end{abstract}


\clearpage
%=========================================================================================================================
\section{Introduction}\label{sec:intro} % 30-APR-2021
In science there is a history of scholarly publishing stings \citep{Faulkes2021}.
Scholars and journalists have submitted nonsensical papers to various venues to expose dysfunctional peer review.
These nonsensical papers submitted can be written by humans \citep[e.g., the \href{https://en.wikipedia.org/wiki/Sokal_affair}{\emph{Sokal Affair}} and][]{Bohannon2013} or computer generated (e.g., \href{http://github.com/strib/scigen}{SCIgen}, \href{https://thatsmathematics.com/mathgen/}{Mathgen}).

Computer programs designed to generate fake papers and sting publishers are also reused by academic tricksters who easily produce the (fake) publications or (fake) citations they desperately need.
As a result, meaningless randomly generated scientific papers end up being served and sometimes sold by various publishers with a prevalence estimated to 4.29 papers every one million papers~\citep{VanNoorden2021,CabanacAndLabbe2021}.
Such papers can be easily spotted by both human and machine; natural language generation tools thus appear to be a cheap and dirty alternative to buying publications from paper mills, which also seems on the rise~\citep{ElseAndVanNoorden2021,Mallapaty2020}.

The major recent advances in language models based on neural networks may sooner or later lead to a new kind of scientific writing.
Incorrigible optimists would consider that automatic translation, writing enhancement, and summarising tools help authors to produce better scientific papers.
Whole books are now generated from thousands of articles used as input~\citep{BetaWriter2019,Day2019,GeneratedBook21}.
But the generative power of modern language models can also be considered a threat to the integrity of the scientific literature. For example, the dangerous nature of the GPT-3 language model \citep{BrownEtAl2020} was discussed extensively \citep{Hutson2021}.

With this in mind, we report observations about a reputable journal along several lines: occurrences of tortured phrases in publications (e.g., `flag to clamor' in lieu of the established `signal to noise'), indication -- if not evidence -- of AI-generated abstracts, as well as questionable texts and images (including reuse from other sources without proper acknowledgement), as well as recent changes in editorial management (including shortened time between reception and acceptance of manuscripts).
Without any definitive proof, we thus provide hints of the rise of a new kind of probably synthetic, nonsensical scientific texts.

The outline of this open call for investigation is as follows.
Section~\ref{sec:tort-phases} reports a set of `tortured phrases' spotted in the literature.
We then focus our study on \emph{Microprocessors and Microsystems}, an Elsevier journal in which they concentrate (Sect.~\ref{sec:micropro}).
We report intriguing irregularities in the editorial timelines of this journal (Sect.~\ref{sec:timelines}).
The presence of synthetic text generated by advanced language model is hypothesised and Sect.~\ref{sec:gptscore} reports the screening of recent publications using an off-the-shelf software detecting synthetic text.
Section~\ref{sec:cases} provides factual evidence of inappropriate and/or poor quality publications.
We discuss possible sources of synthetic papers in Sect.~\ref{sec:discussion} before concluding with a call to the scientific community for further investigation on this matter (Sect.~\ref{sec:conclusion}).

%=========================================================================================================================
\section{Tortured phrases found in published academic articles}\label{sec:tort-phases}
While reviewing recent publications, we encountered an unusual and disappointing phenomenon: well-known and well-established scientific terms were replaced by unconventional phrases.
In a typical case, a word-by-word synonymical substitution is applied to a multi-word term.
We call \emph{tortured phrases} these phrases that are incorrectly used in lieu of well-established ones.
Table~\ref{tab:tortured} shows some tortured phrases that we were able to find in the literature (at first by chance and then by snowballing with already identified terms) and retro-engineer to infer the correct wording that readers would expect.

\begin{table}[h]\centering
    \caption{Tortured phrases we found in the literature along with their usual, correct wording.}\label{tab:tortured}
    \resizebox{\textwidth}{!}{%
        \begin{tabular}{ll}\toprule
            Tortured phrase found in publications                                   & Correct wording expected\\\midrule
            profound neural organization                                            & deep neural network\\
            (fake | counterfeit) neural organization                                & artificial neural network\\
            versatile organization                                                  & mobile network\\
            organization (ambush | assault)                                         & network attack\\
            organization association                                                & network connection\\[6pt]
            (enormous | huge | immense | colossal) information                      & big data\\
            information (stockroom | distribution center)                           & data warehouse\\[6pt]
            (counterfeit | human-made) consciousness                                & artificial intelligence (AI)\\[6pt]
            elite figuring                                                          & high performance computing\\
            haze figuring                                                           & fog/mist/cloud computing\\[6pt]
            designs preparing unit                                                  & graphics processing unit (GPU)\\
            focal preparing unit                                                    & central processing unit (CPU)\\[6pt]
            work process motor                                                      & workflow engine\\[6pt]
            facial acknowledgement                                                  & face recognition\\
            discourse acknowledgement                                               & voice recognition\\[6pt]
            mean square (mistake | blunder)                                         & mean square error\\
            mean (outright | supreme) (mistake | blunder)                           & mean absolute error\\[6pt]
            (motion | flag | indicator | sign | signal) to (clamor | commotion | noise) & signal to noise\\[6pt]
            worldwide parameters                                                    & global parameters\\[6pt]
            (arbitrary | irregular) get right of passage to                         & random access\\
            (arbitrary | irregular) (backwoods | timberland | lush territory)       & random forest\\
            (arbitrary | irregular) esteem                                          & random value\\[6pt]
            subterranean insect (state | province | area | region | settlement)     & ant colony\\
            underground creepy crawly (state | province | area | region | settlement) & ant colony\\[6pt]
            leftover vitality                                                       & remaining energy\\
            territorial normal vitality                                             & local average energy\\
            motor vitality                                                          & kinetic energy\\[6pt]
            (credulous | innocent | gullible) Bayes                                 & naïve Bayes \\[6pt]
            individual computerized collaborator                                    & personal digital assistant (PDA)\\
            \bottomrule
        \end{tabular}%
    }
\end{table}

\begin{figure}[t]\centering
    \fbox{\includegraphics[width=.98\linewidth]{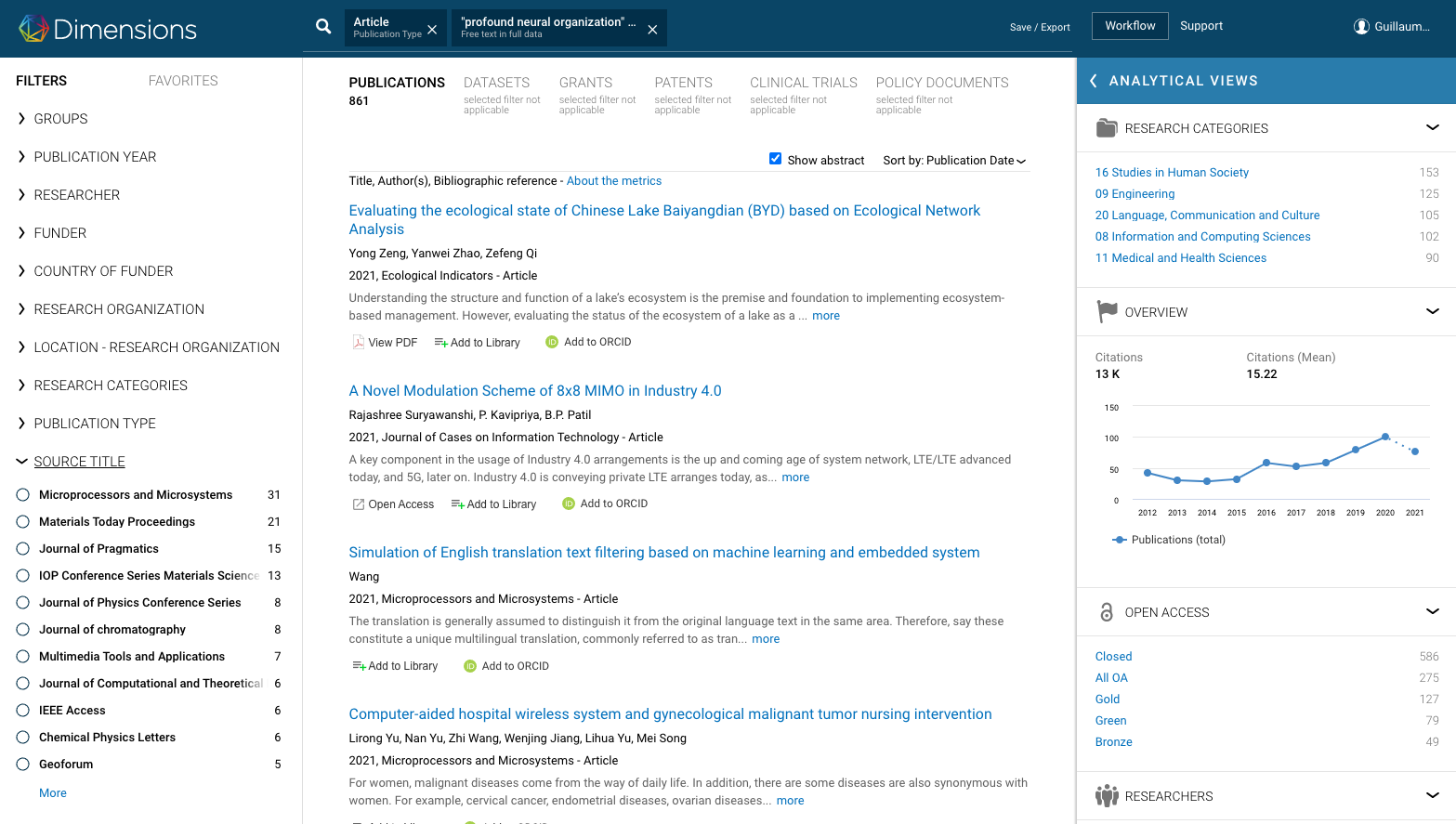}}
    \caption{Published articles retrieved with the Dimensions academic search engine (\url{https://bit.ly/3vm8tAW}).
    The query targets the full-text index with 30 tortured phrases that we listed as of May 25, 2021 (earlier version of Tab.~\ref{tab:tortured}).}\label{fig:microproDimensions}
\end{figure}

On May 25, 2021 we queried the Dimensions academic search engine \citep{HerzogEtAl2020} to retrieve the set of papers containing tortured phrases known at that date (see Fig.~\ref{fig:microproDimensions}).
Note that some tortured phases may be used in a legitimate way (e.g., ‘enormous information’ in certain contexts) and that the full-text indexing performed by Dimensions ignores punctuation.
This may lead to retrieve few articles not using a tortured phrase.
% This indexing leads to the inadvertently retrieval of an article containing a sequence of words that are separated by a punctuation mark (e.g., ‘\ldots{} worldwide. Parameters\ldots’).
Dimensions was chosen for its coverage of the literature that is larger than the Web of Science and Scopus \citep{SinghEtAl2021} and because it is free for scientometric research.\footnote{\url{https://www.dimensions.ai/scientometric-research/}}

The \emph{Microprocessors and Microsystems} journal was ranked first among the venues listed by Dimensions in decreasing number of matching articles (Fig.~\ref{fig:microproDimensions}).
We selected this journal for further investigation in the remainder of this study. 

\clearpage
%=========================================================================================================================
\section{The \emph{Microprocessors and Microsystems} journal}\label{sec:micropro}
% Vol 3 was published by IPC Businness Press Limited (https://doi.org/10.1016/0141-9331(79)90084-X)
Founded in 1976, the \emph{Microprocessors} journal\footnote{\url{https://www.sciencedirect.com/journal/microprocessors}} was quickly renamed \emph{Microprocessors and Microsystems} starting from Volume~3 in 1978.
It is now published by Elsevier\footnote{\url{https://www.sciencedirect.com/journal/microprocessors-and-microsystems}} and classified by Scopus\footnote{\url{https://www.scopus.com/sourceid/15552}} in four subject areas of Computer Science:
\begin{itemize}
    \item Artificial Intelligence
    \item Computer Networks and Communications
    \item Hardware and Architecture
    \item Software
\end{itemize}

Based on Scopus data, \emph{Scimago Journal Ranking} ranked \emph{Microprocessors and Microsystems} in Q3, that is the third quartile for the four subject areas.\footnote{\url{https://www.scimagojr.com/journalsearch.php?q=15552&tip=sid}}
In the latest \emph{Journal Citation Reports} curated by Clarivate Analytics, this journal appears in the \emph{Science Citation Index Expanded} under three categories:
\begin{itemize}
    \item Computer Science, Hardware \& Architecture
    \item Computer Science, Theory \& Methods
    \item Engineering, Electrical \& Electronic
\end{itemize}

As of June 2021, the latest \emph{Journal Citation Reports} entry for \emph{Microprocessors and Microsystems} covered 2017--2019.
The journal published 378 articles with the top~5 contributing countries and organisations in Tab.~\ref{tab:contributors}.
Its \emph{Journal Impact Factor} increased from 0.471 to 1.161 over 2015--2019, that is a 146\% increase over four years.

\begin{table}[h]\centering
    \caption{Top~5 contributing countries and organizations of \emph{Microprocessors and Microsystems} over 2017--2019 as per the \href{https://journalprofile.clarivate.com/jif/home/?journal=MICROPROCESS\%20MICROSY&year=2019&editions=SCIE}{\emph{Journal Citation Reports}} ($N=378$ articles).}\label{tab:contributors}
    \begin{tabular}{lrlr}
    \toprule
    \multicolumn{2}{c}{Countries} & \multicolumn{2}{c}{Organisations}\\\cmidrule(r){1-2}\cmidrule(l){3-4}
    Name & Articles & Name & Articles\\\midrule
    India            & 55 & CNRS, France                              & 22\\
    China (mainland) & 43 & Czech Technical University                &  9\\ 
    France           & 38 & University of Montenegro                  &  9\\
    Germany          & 32 & Technical University of Munich            &  8\\
    USA              & 30 & Universidade Federal do Rio Grande do Sul &  8\\
    Iran             & 28 & Indian Institute of Technology System     &  7\\\bottomrule
    \end{tabular}
\end{table}

\begin{figure}[h]\centering
    \includegraphics[width=\linewidth]{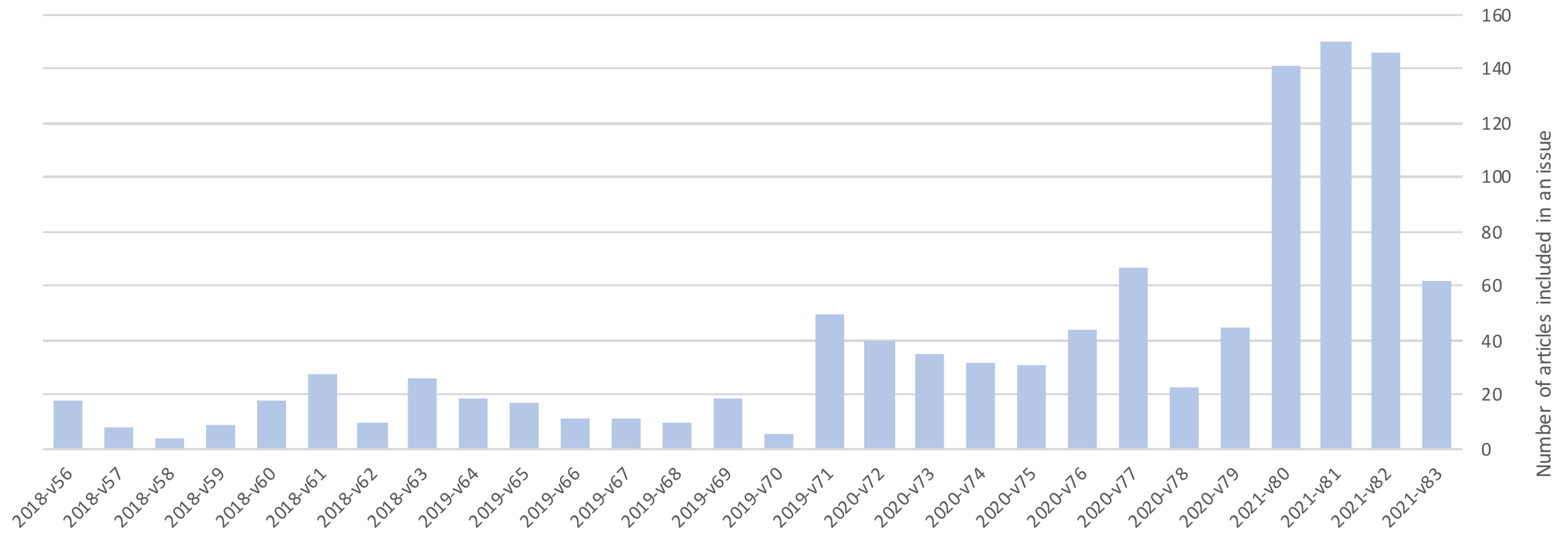}
    \caption{Number of articles included in the volumes 56--83 of \emph{Microprocessors and Microsystems}.}\label{fig:volumes}
\end{figure}

In what follows, we conduct a more in-depth analysis of this venue over the period February 2018 to June 2021 for which we collected data.  Figure~\ref{fig:volumes} shows a radical change in the number of articles published per volumes starting in 2020. 

\emph{Microprocessors and Microsystems} publishes articles with DOIs minted by Crossref \citep{HendricksEtAl2020}.
We queried the Crossref REST API\footnote{\url{https://github.com/CrossRef/rest-api-doc}} to collect the DOIs of papers published in volumes 56--83 (February~2018 to June~2021) of this journal.\footnote{\url{https://www.sciencedirect.com/journal/microprocessors-and-microsystems/issues}}
We used the Elsevier subscription of the University of Toulouse (GC's affiliation) to download each article in fulltext XML via the Elsevier API\footnote{\url{https://dev.elsevier.com}} and extract the following metadata:
\begin{itemize}
    \item Identifiers: Publisher Item Identifier (PII) and Digital Object Identifier (DOI)
    \item Timeline: dates of submission, revision, and acceptance
    \item Publication type (e.g., full-length article, review article, editorial, erratum)
    \item Title
    \item Abstract
    \item Authors' countries
\end{itemize}

We filtered out publication types other than ‘full-length articles’ and removed two articles with a missing acceptance date.
The revision date was missing for 41 articles; we assumed acceptance without revision for these.
We noted that no countries were present in the XML format for 12 articles.
The final dataset contains 1,078 articles (\supplmat).

%=========================================================================================================================
\section{Irregularities of the editorial assessment in \emph{Microprocessors and Microsystems}}\label{sec:timelines}

We use the term `editorial assessment' to denote the time from submission of a manuscript to its acceptance, including: preliminary screening, invitation of reviewers, rounds of peer review, and final decision.
The published metadata for each paper characterises its editorial assessment with three dates: submission, revision, and acceptance.

The analysis of the dates of submission vs dates of acceptance reveals a sudden shortening of editorial assessment for volumes published in 2021.
Most articles were published after an editorial assessment surprisingly short.
Affiliations from China and India were over-represented.
Several blocks of articles shared the same dates of submission and acceptance.
These observations depart from the typical publication output of \emph{Microprocessors and Microsystems} before 2021.

Our call for investigation (Sect.~\ref{sec:conclusion}) invites readers to perform a deeper analysis along the same lines and compare with other reputable journals.

%-------------------------------------------
\subsection{Shortening duration of editorial assessment}
\begin{figure}\centering
    \includegraphics[width=.4\linewidth]{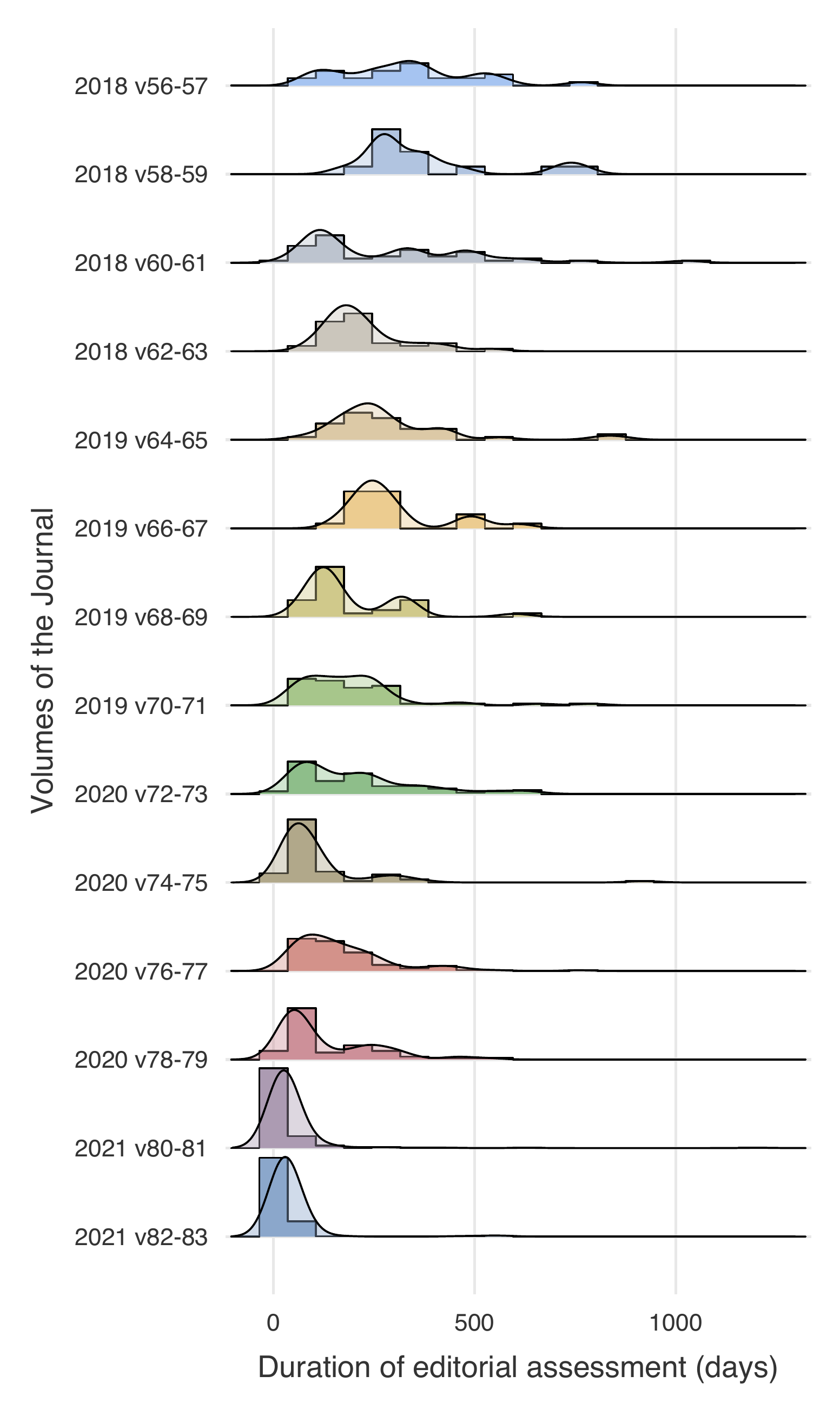}\hfill%
    \includegraphics[width=.5\linewidth]{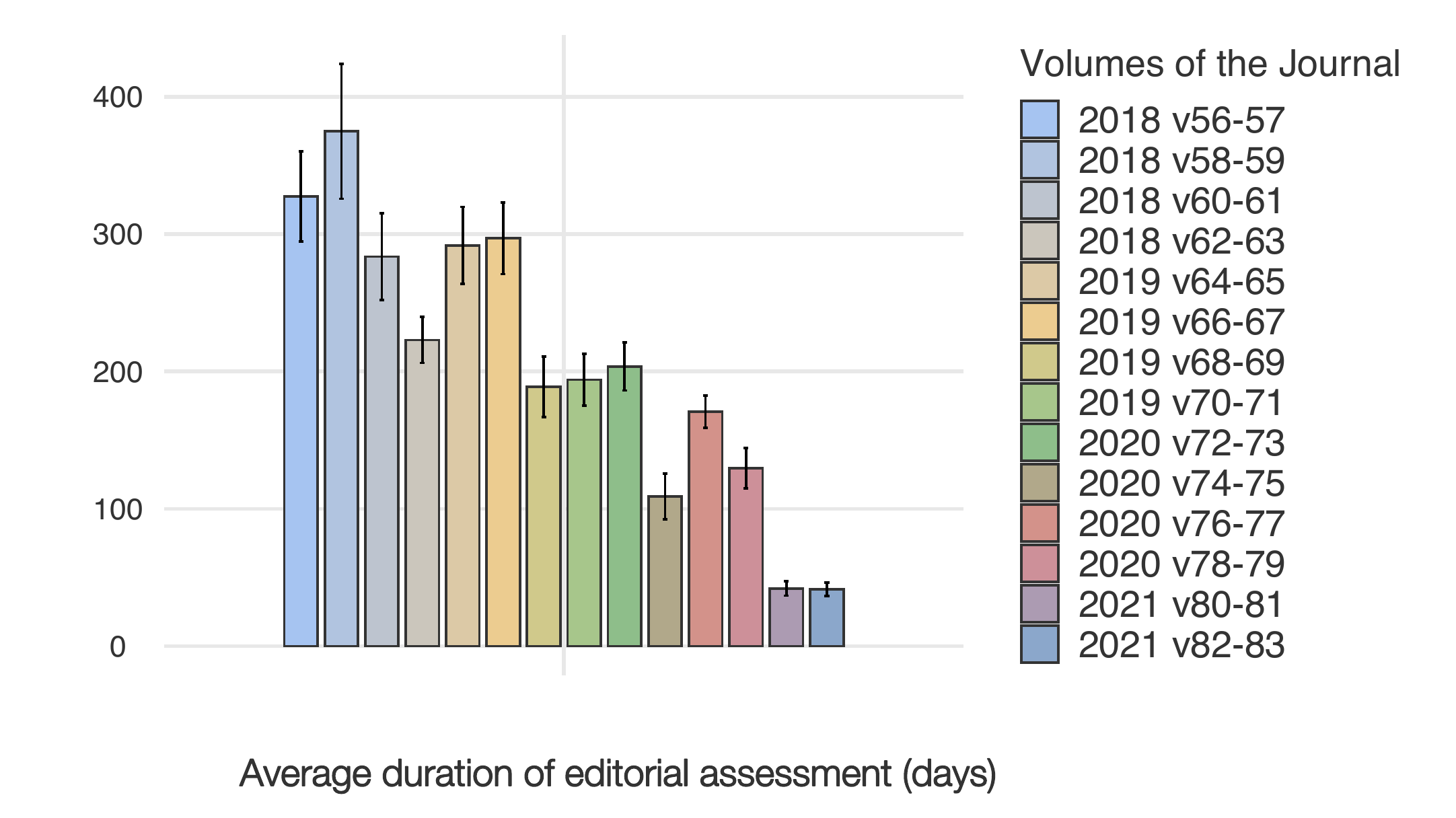}\\
    \includegraphics[width=\linewidth]{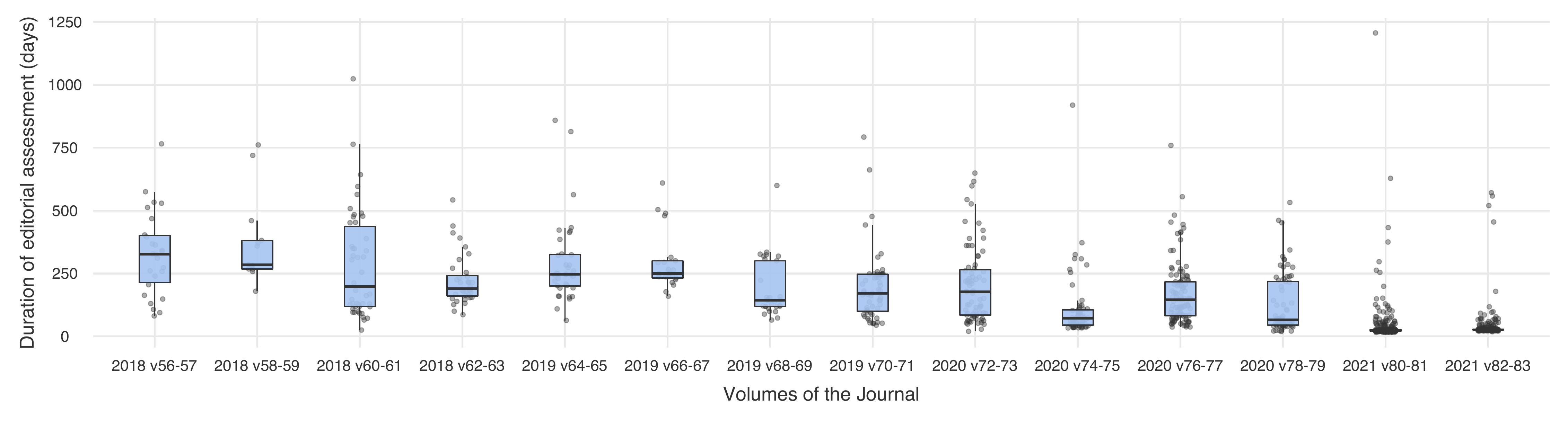}
    \caption{Editorial assessment at \emph{Microprocessors and Microsystems}: duration in days elapsed from submission to acceptance of the 1,078 articles published in volumes 56--83 issued between February 2018 and June 2021. The same data are presented with three complementary visualisations.  The volumes of early 2021 (v80--83) show a 186\% increase in number of accepted papers and an editorial assessment duration divided by 4 (v80--83, $N=499$, $Med=25$ days) compared to the volumes of early 2020 (v74--77, $N=174$, $Med=108$), see Tab.~\ref{tab:statsPR}.}\label{fig:peerReview}
\end{figure}

We noted that shorter processing times (below 40 days) became prevalent, starting from volume 80 of February 2021 (Fig.~\ref{fig:peerReview}).
Statistics on the editorial assessment duration (Tab.~\ref{tab:statsPR}) show a 5-fold decrease in average processing time and a 6-fold decrease in median time when comparing the volumes of 2018--2020 and the volumes of early 2021.

\begin{table}[h]\centering
    \caption{Statistics on the editorial assessment duration (in days) for 3 periods of \emph{Microprocessors and Microsystems}.}\label{tab:statsPR}
    \begin{tabular}{lcS[table-format=3]S[table-format=2]*3{S[table-format=3]}S[table-format=4]}\toprule
    Period        & Volumes & \text{$N$} & \text{Min} & \text{Avg} & \text{StdDev} & \text{Med} & \text{Max}\\\midrule
    2018--2020    & 56--79  & 579 & 19 & 202 & 157 & 161 & 1024\\[3pt]
    Early 2020    & 74--77  & 174 & 33 & 148 & 130 & 108 & 919\\[3pt]
    Early 2021    & 80--83  & 499 & 15 & 42 & 82 & 25 & 1206\\\bottomrule
    \end{tabular}
\end{table}

%-------------------------------------------
\subsection{Quicker editorial assessment and over-representation of some author countries}
Out of 404 papers accepted in less then 30 days after submission, 394 papers (97.5\%) have authors with affiliations in (mainland) China.
Out of 615 papers of which editorial processing time exceeded 40 days, 58 papers (9.5\%) only have authors with affiliations in (mainland) China.
This tenfold imbalance suggests a differentiated processing of papers affiliated to China characterised by shorter peer-review duration.

%-------------------------------------------
\subsection{Blocks of similar editorial timelines}
Skimming through the table of contents, we observed that some papers share identical submission/revision/acceptance dates, which is unusual.
This might suggest editorial overload.
We thus aimed to identify these blocks of articles and the magnitude of this phenomenon.

Given a triple of dates $(x, y, z)$ we define a \emph{block} of papers characterised by this triple as follows: a paper belongs to the block if its submission date is either $x$ or $x + 1$, its revision date is either $y$ or $y + 1$ and its acceptance date is either $z$ or $z + 1$.

\clearpage
We identified 111 (overlapping) blocks consisting of 10 or more papers, and 40 blocks consisting of 20 or more papers (\supplmat).
Let us discuss two blocks whose publications appeared in special issues of the journal:
\begin{itemize}
    \item The block generated by dates (November 22, 2020; December 9, 2020; December 14, 2020) consists of 30 papers:
    \begin{itemize}
        \item 23 belong to the \emph{Special Issue on Embedded Processors},
        \item 2 belong to the \emph{Special Issue on Signal Processing},
        \item 1 belongs to the \emph{Special Issue on Internet of People},
        \item 4 are regular papers.\\[-3pt]
    \end{itemize}

    \item The block generated by dates (November 10, 2020; November 24, 2020; November 30, 2020) consists of 24 papers:
        \begin{itemize}
            \item 16 of which belong to the \emph{Special Issue on Embedded Processors},
            \item 8 are regular papers.
        \end{itemize}
\end{itemize}

These examples show that a single block may contain papers from different special issues as well as regular papers.
The special issues of \emph{Microprocessors and Microsystems} are listed online\footnote{\url{https://www.sciencedirect.com/journal/microprocessors-and-microsystems/special-issues} and \url{https://www.journals.elsevier.com/microprocessors-and-microsystems/special-issues}} with the mention `Edited by' followed by the names of the persons in charge.
A few special issues were introduced with a preface\footnote{e.g., \href{https://doi.org/10.1016/j.micpro.2020.103236}{doi:10.1016/j.micpro.2020.103236} and  \href{https://doi.org/10.1016/j.micpro.2020.103187}{doi:10.1016/j.micpro.2020.103187}} where editors present the topics and review process.
The four special issues featuring papers from the two aforementioned blocks are not mentioned in the list of special issues.
We also failed to find any preface to these special issues.

We were not able to propose a satisfactory explanation to this phenomenon within the bounds of a normal editorial process.

\subsection{Discussion}\label{sec:discussionMicroPro}
The observed shortening time between submission and acceptance may reflect poor or deficient editorial assessment. 
Meanwhile, we noted at least two retractions\footnote{\href{https://doi.org/10.1016/j.micpro.2020.103229}{doi:10.1016/j.micpro.2020.103229} and \href{https://doi.org/10.1016/j.micpro.2017.11.007}{doi:10.1016/j.micpro.2017.11.007}} in \emph{Microprocessors and Microsystems} for text duplication, indicating that the journal responds to integrity concerns at least in some cases.
The closing of these two retraction notices are similar (only difference: the wording ‘severe abuse’ or ‘misuse’) and read as:
\begin{quote}
``As such this article represents a (misuse | severe abuse) of the scientific publishing system.
The scientific community takes a very strong view on this matter and apologies are offered to readers of the journal that this was not detected during the submission process.''
\end{quote}

While the suspected low editorial standards may explain how texts with tortured phrases got published, the process by which those tortured phrases were coined is quite mysterious.
It seems improbable, for any skilled scientist, to use a non-standard terminology to refer to well-known concepts in one's field. 
In addition, when authors are able to cite the literature that uses the standard terminology, it is unexpected that they switch to a tortured version of the terminology in their own manuscripts.

Our hypothesis is that the observed tortured phrases were coined by \emph{misused} natural language processing (NLP) tools: automatic translation, automatic re-writing or even automatic generation of text.
Today, the vast majority of these tools relies on advanced language models.
In the next section we investigate a way to detect the use of such models.

%=========================================================================================================================
\section{Abstracts with high Generative Pre-Training (GPT) detector score}\label{sec:gptscore}

Advanced NLP models are now core building blocks for any natural language-related task: translation, information retrieval, classification, named-entity recognition, text generation, and so on.
Regarding text generation, detectors of synthetic texts have been released.
Automatic detection of computer generated text has already drawn attention in the past, see for example~\citep{LabbeEtAl2016,CabanacAndLabbe2021,DalkilicEtAl2006,Amancio2015}.
This section focuses on one of the most recent detectors.

%-------------------------------------------
\subsection{GPT and the GPT-2 Output Detector}

The OpenAI company has released several advanced language models: Generative Pre-training~\citep[GPT,][]{radford2018}, Generative Pre-trained Transformer~2 \cite[GPT-2,][]{SolaimanEtAl2019b}, and GPT-3 \citep{BrownEtAl2020}.
The generative power of these models has been extensively discussed:
\begin{itemize}
    \item ``Humans find GPT-2 outputs convincing. Our partners at Cornell University surveyed people to assign GPT-2 text a credibility score across model sizes.'' \citep{SolaimanEtAl2019b}\\[-2pt]
 
    \item ``We've seen no strong evidence of misuse so far. While we've seen some discussion around GPT-2's potential to augment high-volume/low-yield operations like spam and phishing, we haven't seen evidence of writing code, documentation, or instances of misuse. We think synthetic text generators have a higher chance of being misused if their outputs become more reliable and coherent. We acknowledge that we cannot be aware of all threats, and that motivated actors can replicate language models without model release.'' \citep{SolaimanEtAl2019b}\\[-2pt]

    \item ``With its apparent ability to artificially read and write, GPT-3 is perhaps different from other forms of AI, in that writing seems more fluid, open-ended, and creative than examples of AI that can beat people in a game or classify an image''~\citep{VenemaEtAl2020}\\[-2pt]

    \item There is anecdotal evidence\footnote{See \url{writemeanabstract.com} and \url{https://twitter.com/DrJHoward/status/1188130869183156231}} that GPT-2 was re-trained on Pubmed abstracts to generate scientific texts \citep{Lang2019}.
\end{itemize}

% We also found a counterexample: Nature paper flagged 99\% with rbase \url{https://www.nature.com/articles/s41586-021-03422-5}
A report from OpenAI discusses the ability of humans to differentiate between genuine texts and texts generated with GPT-2.
It also presents and evaluates different versions of classifiers aiming to detect synthetic text and claims ``Our classifier is able to detect 1.5 billion parameter GPT-2-generated text with approximately 95\% accuracy'' \citep[p.~10]{SolaimanEtAl2019}.
Unfortunately, the report does not provide any clue about precision and recall (i.e., false positive and false negative rates).
Nevertheless, several versions of GPT-2 detectors are provided along with the generators so to flag synthetic texts.
One is based on RoBERTa \citep{LiuEtAl2019} and available as a website called ‘\href{https://huggingface.co/openai-detector}{GPT-2 Output Detector Demo}.’
This detector estimates a `fake' score for a text given as input, reflecting the probability the text was generated. This prediction comes with a caveat: `The results start to get reliable after around 50 tokens.'

How GPT-2 relates to tortured phrases?
Some non-native English authors write in their mother tongue and then translate into English using a translation service, such as Deepl or Google Translate.\footnote{See \url{https://www.deepl.com/translator} and \url{https://translate.google.com}}
We hypothesised that observed tortured phrases would result from advanced language models: either through uncorrected translations or through text generation.
Using a sample of texts in French, we checked the GPT-2 detector score before and after translation into English.
The automatically translated results were clearly marked as `fake.' 
This suggests that the GPT-2 detector flags text generated using GPT-2 \emph{and} synthetic texts from other sources.
If true, the GPT-2 detector may prove useful to flag questionable papers.
This we investigate in the next section.

%-------------------------------------------
\subsection{Datasets for evaluation}

First, we retrieved abstracts for all full-length articles from volumes 80--83 of \emph{Microprocessors and Microsystems} that were processed in less than 30 days.
We thus obtained a set of 389 articles, which we call the \emph{experimental set}.
Given our earlier observations about editorial timelines, it is natural to expect articles from this set to be `probably questionable.'
Table~\ref{tab:lessThan30d} shows a breakdown of the 389 articles by special issue; regular papers are accounted for separately.

Having run the RoBERTa base GPT detector\footnote{\url{https://github.com/openai/gpt-2-output-dataset/tree/master/detector}} against all abstracts of the articles in the experimental set, we observed a prevalence of high GPT detector scores, see Tab.~\ref{tab:rbase}.
Then we proceeded to assemble control sets to pursue the following goals:
\begin{itemize}
    \item Answering the question, ``do abstracts from \emph{Microprocessors and Microsystems} exhibit a higher prevalence of articles with greater GPT detector scores compared to other sets of articles?''
    
    \item Finding possible explanations for prevalence of high GPT detector scores, other than use of GPT. The GPT detector may be sensitive to output of other advanced language models, at the basis of automatic translation or cross-language (self-)plagiarism.
\end{itemize}

Five control sets were created, each consisting of 50 samples except for the last one:
\begin{itemize}
    \item[(A)] The abstracts of 50 most recent (by acceptance date) articles published in volumes 57--79 and processed in 41 days or more. We expected this set to represent ``least concerning articles'' from \emph{Microprocessors and Microsystems}.
    
    \item[(B)] The abstracts of the 50 most recent articles in a selected set of SIAM journals, conditioned that the full text of the article contain terms among: IoT, wireless, sensor, sensors, deep learning, neural network, and neural networks.
    
    \item[(C)] The same set of abstracts as (B), but translated to Chinese and then back to English using Google Translate. 
    
    \item[(D)] 50 Chinese-language abstracts from {\it Wireless Internet Technology}\footnote{\url{https://wap.cnki.net/touch/web/Journal/Index/WXHK.html}} translated to English using Google Translate. The abstracts were selected at random from volumes 1/2021 and 2/2021.  Before sampling, we excluded 3 abstracts from Volume 1/2021 that appeared to be advertisements of other journals.
    
    \item[(E)] 139,236 abstracts from randomly-selected articles published in 2021 by Elsevier.  We retrieved this sample from the Web of Science on May 21, 2021 with  query {\smaller\textsf{PY=2021 AND PUBL="Elsevier" AND DT="Article" AND LA="English"}} run on the \emph{Science}, \emph{Social Science}, \emph{Art and Humanities}, and \emph{Emerging Sources} citation indexes.
\end{itemize}

The control set (A) was chosen to represent the content of {\it Microprocessors and Microsystems} before the apparent change in the journal's operation mode. 

The control set (B) was expected to represent high-quality, well-written, and thoroughly proofread articles. We infer these traits from the reputation of the society that publishes the journals. In addition, by selecting articles with certain terms we aim to ensure similarity with the experimental set by topic. 

The control sets (C) and (D) were created in order to emulate a situation in which an English-language paper is prepared using automated translation from some other language.

The control set (E) was created to reflect a large proportion of the articles Elsevier published in early 2021 irrespective of the journals and scientific fields.

\begin{table}[h]\centering
    \caption{Articles of the Experimental set: breakdown by Special Issue.}\label{tab:lessThan30d}
    \begin{tabular}{lrrr}\toprule
        Headings                                & \multicolumn{2}{c}{Articles vol. 80--83} & Share\\\cmidrule{2-3}
                                                & editorial assessment <30d (Exp)        &  all articles & (\%) \\\midrule
        Special issue on Signal Processing      &         155	& 176	                 & 88.1\\
        Special issue on Internet of People     &          98	& 102	                 & 96.1\\
        Special issue on Embedded Processors    &          74	& 84	                 & 88.1\\
        Regular Papers	                        &          49	& 83	                 & 59.0\\
        Special issue on AI-SIGNAL PROCESSING	&          12	& 18	                 & 66.7\\
        Special issue on CyberSECHARD2019	    &           1	& 4	                     & 25.0\\\midrule
        Grand Total                             &         389   & 467                    & 83.3\\\bottomrule
    \end{tabular}
\end{table}

%-------------------------------------------
\subsection{Results and analysis}

The evaluation of abstracts from experimental and control sets against the RoBERTa base GPT detector is given in Tab.~\ref{tab:rbase}.

\begin{table}[h]\centering
\caption{Distribution (in \%) of GPT detection scores by RoBERTa base.  Rounded values may add up to greater than 100.0.}\label{tab:rbase}
    \begin{tabular}{c*6{S[round-mode=places,round-precision=1,round-integer-to-decimal]}} \toprule
        Score & \text{Experiment Set} & \text{Control A} & \text{Control B} & \text{Control C} & \text{Control D} & \text{Control E}\\
                  & \text{($N=389$)} & \text{($N=50$)} & \text{($N=50$)} & \text{($N=50$)} & \text{($N=50$)} & \text{($N=139,236$)}\\\midrule
        $[0.0, 0.1[$ & 8.5 & 78 & 90 & 56 & 28 & 89.9\\
        $[0.1, 0.2[$ & 1.5 & 6 & 6 & 4 & 12 & 2\\
        $[0.2, 0.3[$ & 0.8 & 0 & 2 & 2 & 8 & 1.1\\
        $[0.3, 0.4[$ & 0.5 & 0 & 0 & 10 & 4 & 0.8\\
        $[0.4, 0.5[$ & 0.5 & 0 & 0 & 0 & 0 & 0.7\\
        $[0.5, 0.6[$ & 1.5 & 0 & 0 & 2 & 2 & 0.6\\
        $[0.6, 0.7[$ & 1.5 & 0 & 2 & 2 & 0 & 0.6\\
        $[0.7, 0.8[$ & 2.1 & 0 & 0 & 2 & 6 & 0.6\\
        $[0.8, 0.9[$ & 2.1 & 0 & 0 & 2 & 4 & 0.8\\
        $[0.9, 1.0]$ & 81 & 16 & 0 & 20 & 36 & 3\\\midrule
        Sum          & 100 & 100 & 100 & 100 & 100 & 100\\
\bottomrule
    \end{tabular}
\end{table}

For the further analysis it is important to stress its main limitation: we assume that GPT scores in each considered case are sampled independently from a distribution related to the case. We consider this assumption reasonable within our setup.

Each of the six samples yields an empirical distribution function which we denote by $F_{exp}$, $F_A$, $F_B$, $F_C$, $F_D$ and $F_E$ for the experimental set and control sets A, B, C, D and E, respectively. A confidence band is then constructed around each empirical distribution function using the Dvoretzky--Kiefer--Wolfowitz inequality, see \citep{DvoretzkyEtAl1956} for the original work and \citep{Massart1990} for the inequality with sharp constants; an exposition is also available in Section~II of \citep{LearnedMillerDeStefano2008}.

\begin{equation}\label{eq:dkw}
    \Pr \left( \sup\limits_x |F_{emp}(x) - F(x)| > \varepsilon \right) \leq 
    2\exp(-2N \varepsilon^2),
\end{equation}
where $N$ is the sample size, $F_{emp}$ is the empirical distribution function, $F$ is the cumulative function of the underlying distribution. Adjusting the value of $\varepsilon$ will allow to control the right-hand side of~\eqref{eq:dkw} as follows:
\begin{equation*}
    \alpha = 2\exp(-2N \varepsilon^2) \quad \Leftrightarrow \quad
    \varepsilon = \sqrt{\frac{\ln (2 / \alpha)}{2N}}.
\end{equation*}

We chose $\alpha = \frac{1}{120}$ so that with 95\% ($0.95 = 1 - 6 \cdot \frac{1}{120}$) confidence all cumulative distribution functions for the 6 considered cases lie within their respective confidence bands. This approach is designed to account for multiple comparisons of the experimental case with control cases. 
For the distribution sizes at hand, i.e. $N = 389$, $N = 50$ and $N = 139,236$, the half-width of the respective confidence bands are $\varepsilon \approx 0.084$, $\varepsilon \approx 0.234$ and $\varepsilon \approx 0.004$, respectively.

\begin{figure}\centering
    \subfloat[Experiment vs Control A]{\includegraphics[width=.48\linewidth]{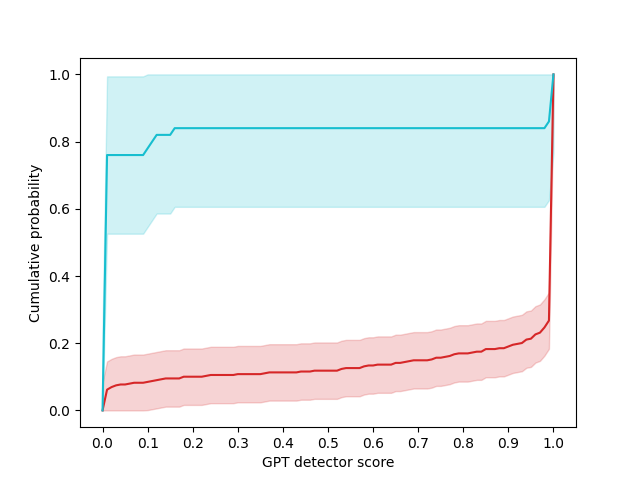}}\hfill 
    \subfloat[Experiment vs Control B]{\includegraphics[width=.48\linewidth]{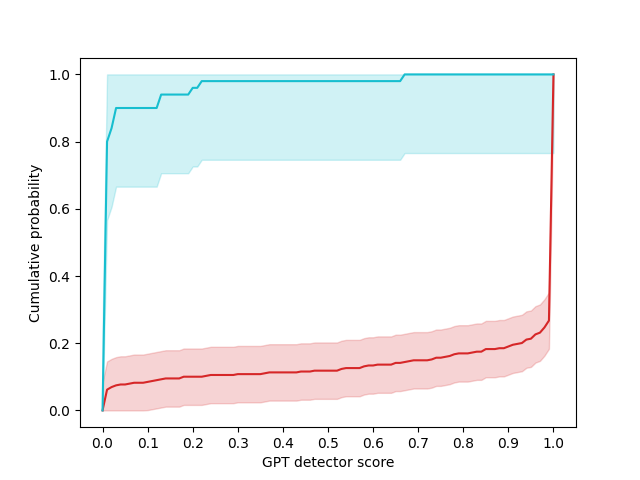}}\\[-6pt]
    \subfloat[Experiment vs Control C]{\includegraphics[width=.48\linewidth]{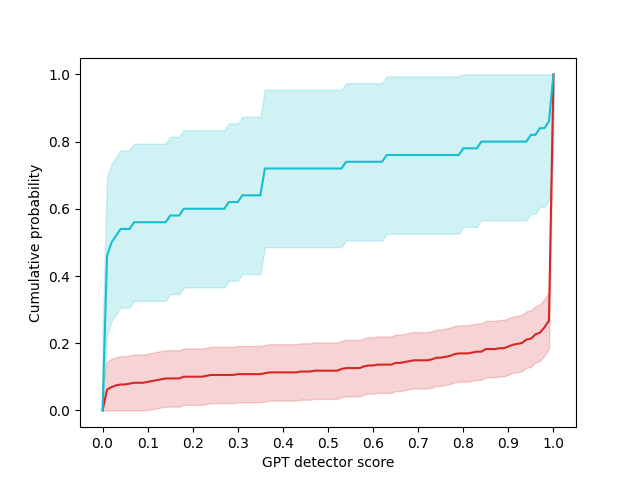}}\hfill
    \subfloat[Experiment vs Control D]{\includegraphics[width=.48\linewidth]{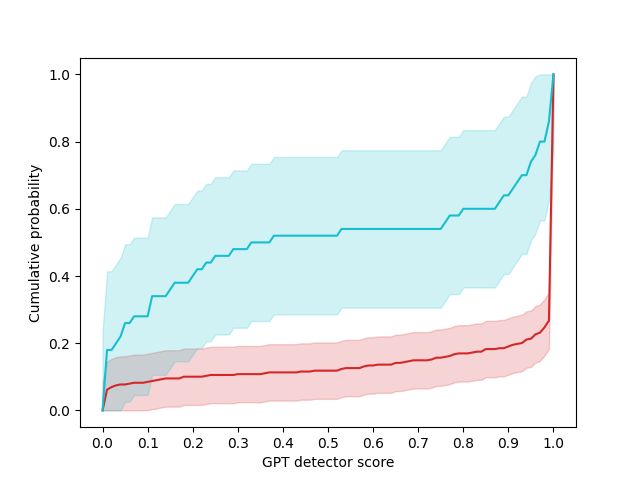}}\\[-3pt]
    \subfloat[Experiment vs Control E]{\includegraphics[width=.48\linewidth]{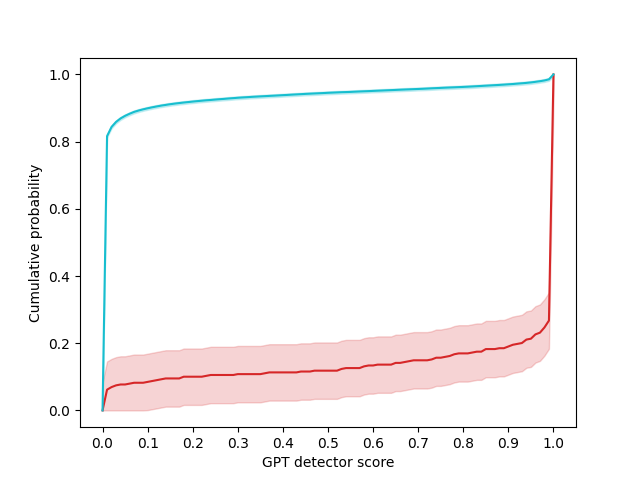}}
    \caption{Cumulative distribution functions for the Experiment set (red line) vs Control sets A--E (blue line) with confidence bands (Tab.~\ref{tab:rbase}).}\label{fig:CDFplots}
\end{figure}

In Fig.~\ref{fig:CDFplots} the confidence band for the experimental case is plotted against each confidence band for control cases. We hereby see that any of the cut-offs $0.3$, $0.4$, \ldots, $0.9$ distinguishes the experimental case from all control cases in the sense that the scores in the experimental case occur above the selected cut-off significantly more often than the scores for the control cases do so. With the exception of Control~D (abstracts from a Chinese journal translated to English), the same holds for the cut-offs $0.1$ and $0.2$. Under the design of our comparison, it is undecided whether the cut-offs $0.1$ and $0.2$ draw distinctions between the experimental set and the control set D; we hypothesise that the small size of set~D does not provide sufficient power for comparison.

Table~\ref{tab:gptEls2021} reveals that several journals published articles with abstracts having a 70\% or higher GPT detector score.
The 70\% threshold was selected because it belongs to the flat part of the cumulative distribution function of Control set~E. 
Let us stress that the concentration of articles with high GPT scores is outstanding in \emph{Microprocessors and Microsystems} with 72.1\% compared to 13.6\% maximum in the other journals tabulated.
The column `Number of articles' shows that many journals published papers with abstracts featuring a high GPT score.
While a high GPT score for an abstract does not necessarily indicate flaws in an individual paper, high concentrations of such articles in certain venues invites a further assessment of this phenomenon.

\begin{table}[h]\centering
    \caption{Elsevier journals from Control~E with 25+ articles published in 2021 whose GPT detector score for abstracts is 70\% or higher. The journal under investigation in this study, \emph{Microprocessors and Microsystems}, has 75 articles with a 70\% or higher GTP score for abstracts ($\mathit{Avg} = 98.6$). These 75 articles represent 72.1\% of all articles published in this journal that are in Control~E.}\label{tab:gptEls2021}
    \resizebox{\textwidth}{!}{%
        \begin{tabular}{>{\itshape}lS[round-mode=places, round-precision=1, round-integer-to-decimal]SS[round-mode=places, round-precision=1, round-integer-to-decimal]}\toprule
            \upshape{Journal} & \multicolumn{2}{c}{Article abstracts with GPT detector score $\geq$ 70\%} & \text{Total articles}\\\cmidrule(lr){2-3}
                    &  \text{Average GPT detector score (\%)} & \text{Number of articles} & \text{in journal (\%)}\\\midrule
            J. Alloy. Compd. & 92.7 & 104 & 5.6\\
            Sci. Total Environ. & 93 & 81 & 3.3\\
            J. Clean Prod. & 92.5 & 76 & 4.4\\
            \rowcolor{Cornsilk}
            Microprocess. Microsyst. & 98.6 & 75 & 72.1\\
            J. Mol. Struct. & 93.3 & 73 & 9.5\\
            Chem. Eng. J. & 90.8 & 68 & 4.3\\
            Appl. Surf. Sci. & 93.8 & 64 & 5.3\\
            Mater. Lett. & 93.5 & 62 & 9.8\\
            Ceram. Int. & 91.5 & 61 & 4.9\\
            Sens. Actuator B-Chem. & 92.3 & 48 & 7.5\\
            Int. J. Hydrog. Energy & 91.4 & 47 & 5\\
            Chemosphere & 93.6 & 47 & 3.6\\
            J. Colloid Interface Sci. & 94.6 & 46 & 5.4\\
            J. Hazard. Mater. & 92.9 & 45 & 3.7\\
            J. Mol. Liq. & 91.2 & 45 & 6.7\\
            Biochem. Biophys. Res. Commun. & 93.4 & 42 & 7.9\\
            Renew. Energy & 93.1 & 39 & 5.4\\
            J. Comput. Appl. Math. & 92.7 & 39 & 13.6\\
            Fuel & 91.6 & 39 & 3.6\\
            Constr. Build. Mater. & 93.5 & 39 & 3.5\\
            Spectroc. Acta Pt. A & 91.5 & 38 & 7.3\\
            Energy & 93.9 & 34 & 4.1\\
            Bioresour. Technol. & 92 & 34 & 6.5\\
            Food Chem. & 90.8 & 33 & 2.6\\
            Powder Technol. & 94.5 & 32 & 8.7\\
            Measurement & 94 & 31 & 6.1\\
            Carbohydr. Polym. & 90.1 & 31 & 5.3\\
            J. Differ. Equ. & 89.1 & 27 & 11.2\\
            Electrochim. Acta & 89.8 & 27 & 5.5\\
            Alex. Eng. J. & 92.6 & 26 & 9.8\\
            Opt. Laser Technol. & 90.7 & 25 & 8.7\\
            Neurosci. Lett. & 94.6 & 25 & 11.3\\
            J. Math. Anal. Appl. & 89.3 & 25 & 7.5\\
            Ecotox. Environ. Safe. & 92.2 & 25 & 4.4\\
            Environ. Pollut. & 91.2 & 25 & 3.4\\
            Carbon & 92.8 & 25 & 6.6\\
            \bottomrule
        \end{tabular}}
\end{table}

The concentration of abstracts with a high GPT detector score in \emph{Microprocessors and Microsystems} (Experimental Set) is intriguing.
Nonetheless, texts flagged as synthetic by the GPT detector might be scientifically sound.
We visually examined several publications from this journal to go beyond automatic screening.
The next section reports critical flaws we found in several papers, including nonsensical text featuring tortured phrases, plagiarised text, and image theft.
We believe these publications should be considered for retraction as they ``represent a severe abuse of the scientific publishing system'', as quoted in the retraction notices reproduced in Sect.~\ref{sec:discussionMicroPro}.

%=========================================================================================================================
\section{Critical flaws found in questionable and problematic publications: individual cases}\label{sec:cases}

All the above quantitative observations suggest that certain editorial processes in several venues were (and might still be) arranged in a non-conventional manner. In order to test this hypothesis, we analysed several individual papers from the journal {\it Microprocessors and Microsystems}. For each case presented in this section, we expose various flaws that, in our opinion, are unacceptable in published scientific literature.
Our observations include:
\begin{itemize}
    \item reuse of text and / or images without acknowledgement;
    \item references to non-existing literature;
    \item references to non-existing internal entities of the paper (e.g., theorems and variables in formulas);
    \item sentences for which we failed to infer any meaning.
\end{itemize}

Excerpts from each case are reported with a score computed by the \href{https://huggingface.co/openai-detector/}{GPT-2 Output Detector} for which ``The results start to get reliable after around 50 tokens.''
We searched \href{https://www.google.com/imghp}{Google Images} --- either via the `Search by image' feature or by typing in characteristic keywords --- for potential earlier occurrences of selected images that appeared most suspect to us (e.g., irrelevant, of poor visual quality) in the papers we inspected.
Note that we did not perform this image screening systematically.
As of July 8, 2021 there were no citations for the six cases except for Case~5 and Case~6 with one citation each.

While mentioning individual papers, we explicitly refrain from including them in our list of references not to distort the scholarly record.
None of the cases we discuss in the following sections had been reported to PubPeer \citep{BarbourAndStell2020}.
We posted a PubPeer comment for each to trigger discussions.

Our purpose is not to blame individual authors but to trigger the necessary investigation to be conducted by editors and publishers.

\newcommand{\onecase}[2]{$\rhd$ \emph{#1}, #2\\}

%-------------------------------------------
\subsection{Case 1: Unacknowledged (mis)use of a water leak detector description from elsewhere}

\onecase{Real time monitoring of medical images and nursing intervention after heart valve replacement}{Published in volume 82 of April 2021, \href{https://doi.org/10.1016/j.micpro.2020.103766}{doi:10.1016/j.micpro.2020.103766}.}

The English is hard to understand and the meaning of some sentences is quite difficult to infer. For example, the section \emph{literature survey} starting on the first page reads as follows: 
\faketbox{Case 1, \emph{literature survey} -- GPT detector score: 99.98\%}
{
    A \textbf{pamphlet of sickness} or harmed heart valves, ailment, or passing is one of the world’s significant reasons. \textbf{Accessible medicines} for patients with a heart valve \textbf{are abused}; however, to fix the valve because the fix is incredible, \textbf{it have} to supplant a heart valve in the most genuine cases.
}

%Or 
% section Prosthetic heart valve surgery for patients start with :
% At present, need to make the impacts of utilizing various sorts of the fixing material to fix the absence of present moment and long haul mitral valve, and an appropriate valve cooptation valves, those that require tissue extra flyer assessment has.

\begin{figure}[h]
    \subfloat[Figure in Case 1 and its caption.]{
    %taken from \href{https://doi.org/10.1016/j.micpro.2020.103766}{doi:10.1016/j.micpro.2020.103766}]{
    \includegraphics[width=.42\linewidth]{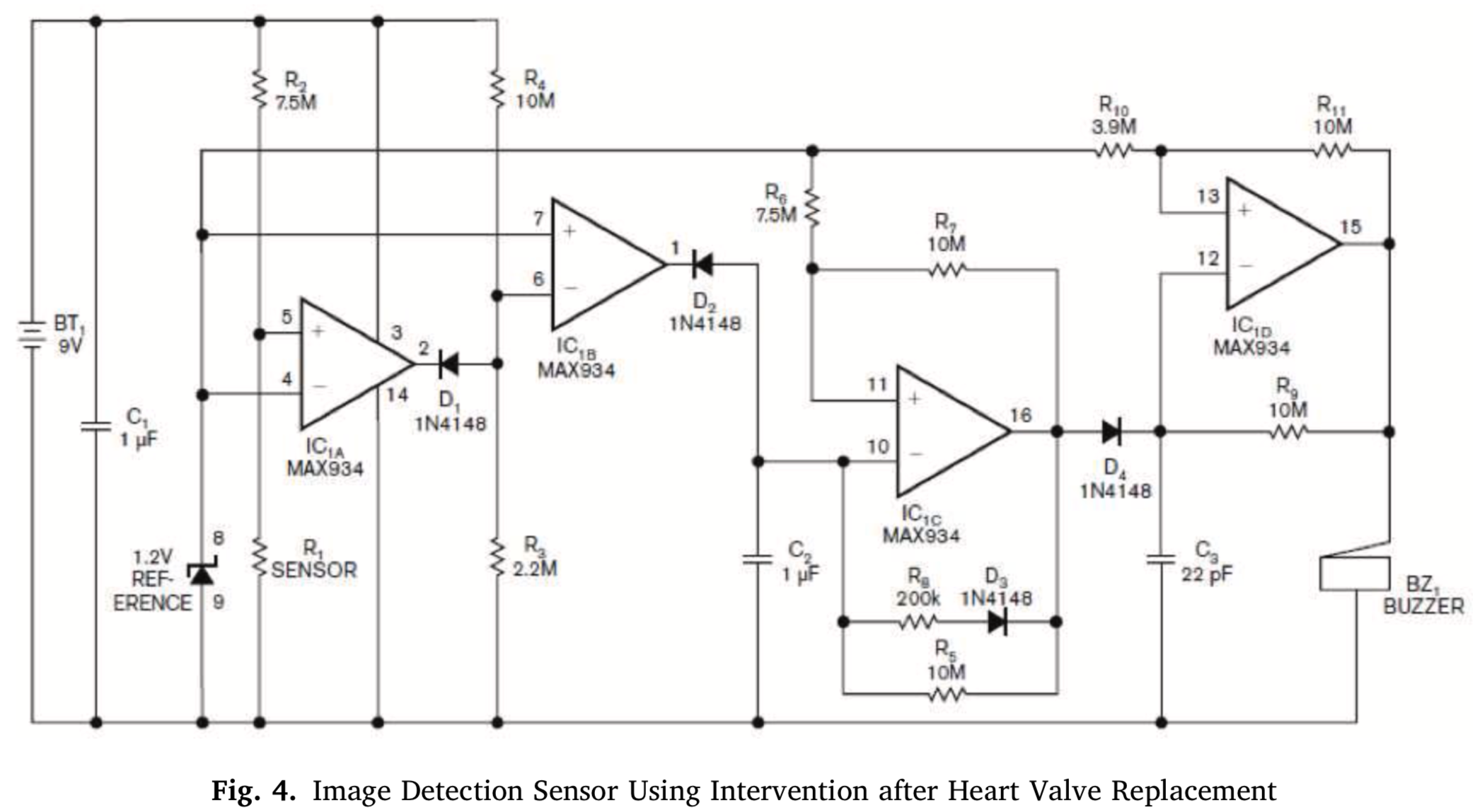}}
    \hfill
    \subfloat[Original figure and its caption.]{
    \includegraphics[width=.42\linewidth]{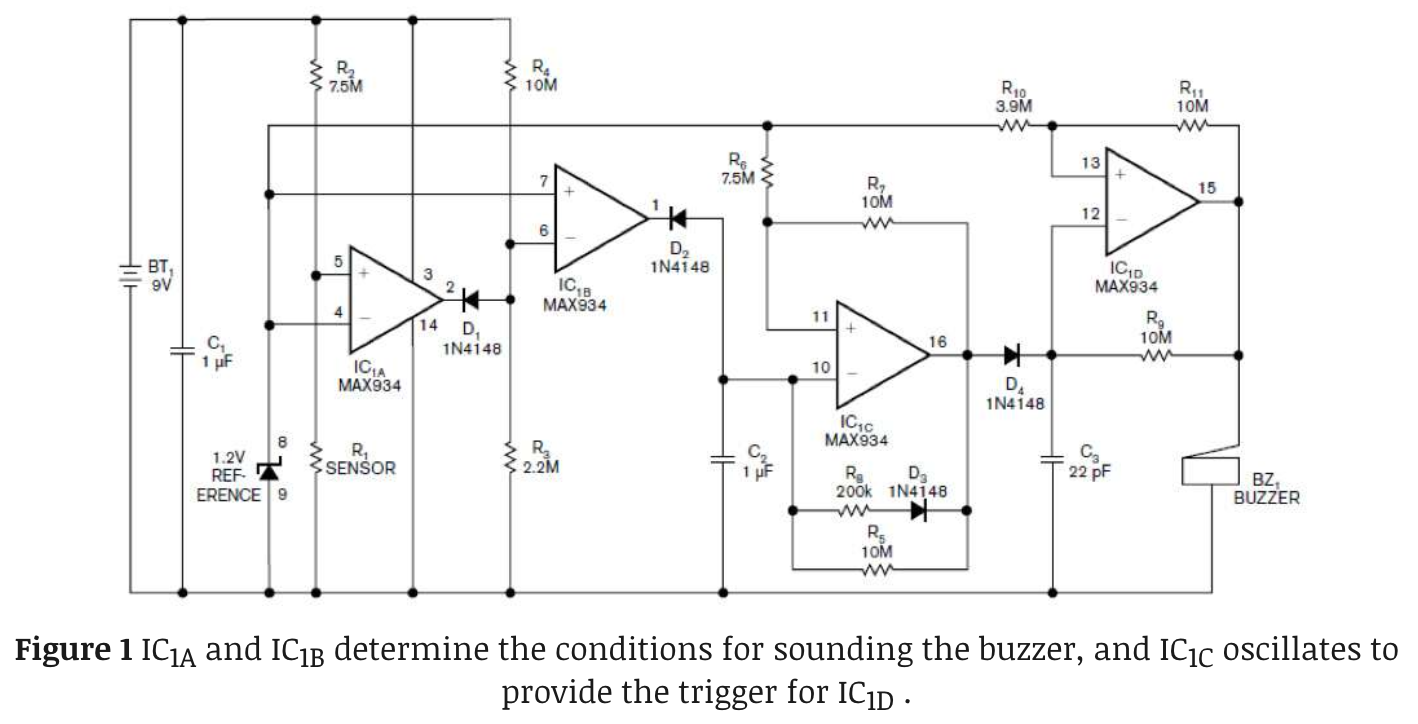}}
    \caption{Case~1 (a) reusing without acknowledgement an original image (b) taken from \url{https://www.edn.com/water-leak-detector-uses-9v-batteries/}.}\label{fig:case1}
\end{figure}

Figure~\ref{fig:case1} shows a figure and its caption that are mostly irrelevant to each other.
In fact, it appears that both the figure and the corresponding paragraph come from \url{https://www.edn.com/water-leak-detector-uses-9v-batteries/}, where a logical diagram for a water leak detector is presented.
The original text has been heavily modified making it hard to understand.\\[6pt]

\noindent
\begin{minipage}{0.45\textwidth}
    \faketbox{Case 1 -- GPT detector score: 88.22\%}
    {
    Fig. 4 shows this design, Image Detection Sensor built-in 1.2V reference Maxim, circuit MAX934, use integrated the four comparators of ultra-low power consumption. {\bf Power consumption of the chip is about 6 $\mu$A}. IC1A, R1, and R2 provides a water leak detection. R1 is water detector may be a two bare copper wire wound around the sponge. {\bf R1 is, because the sponge is dry, when the left output of IC1A is high, you have a high impedance}. Circuit detects a water leakage, the value of R1 is less than one hundred, reduced to several kilo-ohms, and it is the low output of the force IC1A. Through D1, it will make the output of high IC1B.
    }
\end{minipage}
\hfill
\begin{minipage}{0.5\textwidth}
    \truetbox{Case 1's source -- GPT detector score: 3.53\%}
    {
    The design uses Maxim Integrated Circuits'  MAX934, an ultra-low-power quad comparator with a built-in 1.2V reference. {\bf The chip uses about 6 $\mu$A}. IC1A, R1, and R2 provide water-leakage detection. R1 is the water probe, which can be two bare copper wires wrapped in a sponge. {\bf R1 has high impedance when the sponge is dry, so IC1A's output stays high.} Once the circuit detects the water leak, R1's value decreases to less than a few hundred kilohms, which forces IC1A's output low. Through D1, it makes the output of IC1B high.
    }
\end{minipage}

Moreover, Fig.~3 of Case~1 (not shown here) is identical to Fig.~7 (not shown here, caption: `Magnetic resonance imaging in prosthetic heart valves\ldots') published in a 2015 article \href{https://doi.org/10.1161/circimaging.115.003703}{doi:10.1161/circimaging.115.003703} with no visible acknowledgement to the original source of the image.

%-------------------------------------------
\subsection{Case 2: Image reuse}

\onecase{Case 2.1: Computer aided medical system design and clinical nursing intervention for infantile pancreatitis}{Published in volume 81 of March 2021, \href{https://doi.org/10.1016/j.micpro.2020.103761}{doi:10.1016/j.micpro.2020.\allowbreak{}103761}.}
\onecase{Case 2.2: Big Data Prediction of Sports Injury Based on Random Forest Algorithm and Computer Simulation}{Not included in a volume, online since January 2021, \href{https://doi.org/10.1016/j.micpro.2021.104002}{doi:10.1016/\allowbreak{}j.micpro.2021.104002}.}

Case~2.2 contains the following tortured phrases (Tab.~\ref{tab:tortured}): \emph{irregular timberland} (in lieu of \emph{random forest}) and \emph{innocent Bayes} (in lieu of \emph{naïve Bayes}).

These two papers share a common image (without any mention to each other).  This figure features an unexpected Spanish-language annotation on one of the blocks. It has been obtained by cropping of the figure~5.5 on page~136) of:\\[-24pt]
\begin{quote}
    \item A.~Ram\'\i{}rez Agundis, \emph{Dise\~no y experimentaci\'on de un cuantizador vectorial hardware basado en redes neuronales para un sistema de codificaci\'on de video}, Doctoral thesis, Politecnica de Valencia, 2008. \href{https://doi.org/10.4995/Thesis/10251/3444}{doi:10.4995/Thesis/10251/3444}
\end{quote}

\begin{figure}[h]
    \begin{center}
        \subfloat[Original image and caption in Spanish]{\includegraphics[width=.48\linewidth]{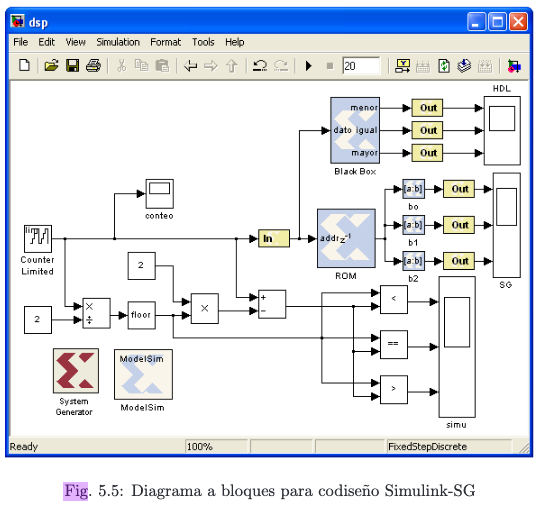}}
    \end{center}
    \begin{center}
        \subfloat[Case 2.1 image and caption]{\includegraphics[width=.48\linewidth]{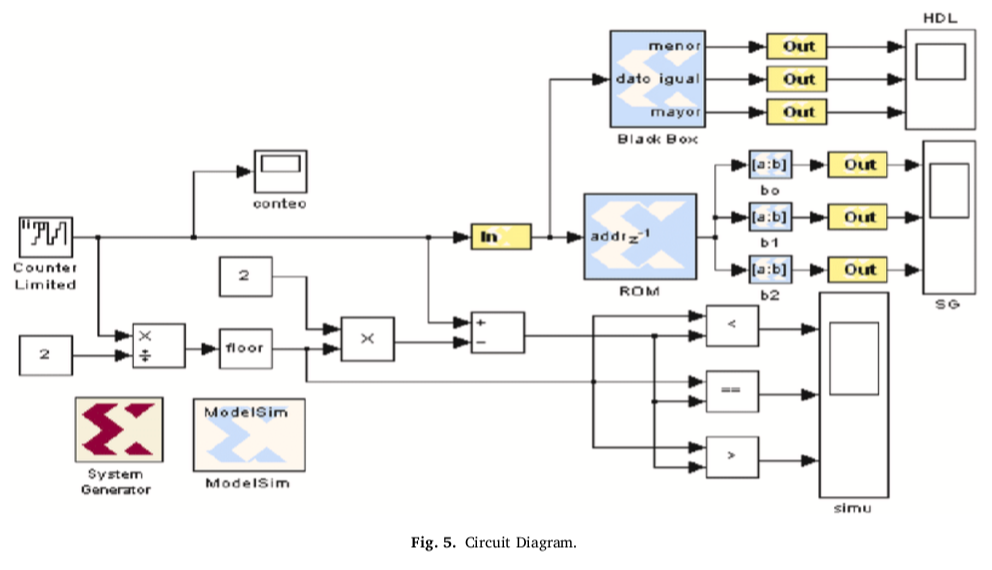}}
        \hfill
        \subfloat[Case 2.2 image and caption]{\includegraphics[width=.48\linewidth]{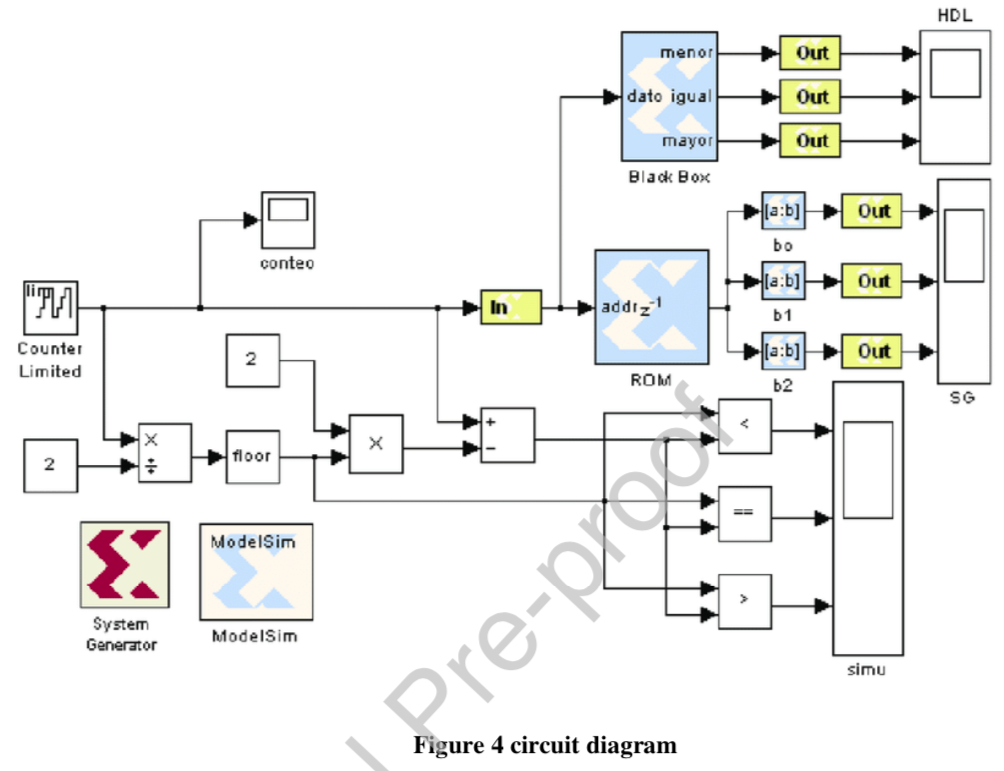}}
    \end{center}
    \caption{Original image from a doctoral thesis~(a) and its cropped versions in Case~2.1~(b) and Case~2.2~(c).}\label{fig:case2}
\end{figure}

The description of the image from each of the two questionable papers, as well as the description of the original image from the thesis by Ram\'\i{}rez Agundis are below. We also reproduce the images themselves in Fig.~\ref{fig:case2}. A notable feature is that the description of the image in Case~2.2 has a relatively low GPT detector score.

\truetbox{Description of the original image -- GPT detector score 0.02\%}{
Como es bien sabido, Simulink es un entorno gr\'afico que permite el dise\~no y simulaci\'on de modelos usando una metodolog\'\i{}a basada en diagramas de bloques.
Proporciona adem\'as bibliotecas de bloques para las tareas comunes de procesamiento en muy diversas \'areas y, loque es importante para el codise\~no, crear nuevas bibliotecas e incorporar otras proporcionadas por terceras partes, en este caso las que suministra el fabricante de los dispositivos FPGA (por lo que se refiere a Xilinx, a trav\'es de System Generator) y las del fabricante de la placa donde est\'a alojada la FPGA. La Fig.~5.5 muestra un ejemplo que incluye bloques de los tres tipos.}

\faketbox{Description of the image in Case 2.1 -- GPT detector score 99.98\%}{
Fig. 5 shown as if all is considered to be completed, after the hunter and viral contamination, T-gloomy period based on the average cycle-specific clinical course staged the febrile period, step-down stage, hypourinary stage diuretic phase. It can be improved—an essential part of all stages of kidney contribution.
}

\faketbox{Description of the image in Case 2.2 -- GPT detector score 36.75\%}{
Figure 4 the following stage expects us to change from likelihood to chances. The pre-test changes can be determined utilizing the recently sketched out condition (see area "Relative dangers and chances proportions") or determined utilizing the pre-test likelihood.
}

Additionally, we show other samples of questionable text in Cases 2.1 and 2.2, with some irregularities highlighted.

\faketbox{Case 2.1, \emph{related work} -- GPT detector score 99.98\%}{
Gadget embedded clinical tools are typically coupled to locate, visualize and test the disease [1]. With such a device, including busy period and \textbf{most manifested people: meters above sea level, cardiovascular screen, screen blood glucose}, the ECG (Electro Cardio Gram), X-ray imaging, the MRI (Magnetic Resonance Imaging). CT (Computed Tomography) and PET (Positron Emission Tomography). (After the elimination of human formalism) the installation size and cost associated with clinical applications increased demand for corrective fully computerized gadgets due to the \textbf{dynamic need to squeeze it} is to promote the development control framework.
}

\faketbox{Case 2.2, \emph{conclusion} -- GPT detector score 99.95\%}{
\textbf{Competitor's exhibition to anticipate, recommend the sequential request of elective forecast techniques.} The proposed innovation is an altered form of the group of the move learning model. \textbf{Utilizing many arranged history neural organizations}, searching the principle focus with the proper boundaries, and spotting the best model with at least wellness work in the initial set. \textbf{To assess the proposed expectation's greatness}, competitors' informational execution index around the globe has been utilized.
}

%...-------------------------------------------
\subsection{Case 3: \emph{Circuits Today} heart rate monitor presented as something else}

\onecase{Computer aided intelligent medical system and nursing of breast surgery infection}{Published in volume 81 of March 2021, \href{https://doi.org/10.1016/j.micpro.2020.103769}{doi:10.1016/j.micpro.2020.103769}.}

Figure~2 in Case~3 contains a clear indication of its source (\href{https://www.circuitstoday.com}{www.circuitstoday.com}) and is most probably reused from \url{https://todayscircuits.wordpress.com/2014/06/26/tc-heart-rate-monitor-using-8051/}. It seems to picture a heart rate monitor
while the reference cited in the caption (number [13] in its reference section) is about \emph{Channel attention module with multiscale grid average pooling for breast cancer segmentation in an ultrasound image}. The reference to \url{https://www.circuitstoday.com/heart-rate-monitor-using-8051}, is consistent with older versions of that page (namely, those before June 2, 2016) --- see \url{https://web.archive.org/web/*/https://www.circuitstoday.com/heart-rate-monitor-using-8051}.\\

The \emph{breast surgery dataset} section contains the following text:
\faketbox{Case 3, \emph{breast surgery dataset} -- GPT detector score 99.87\%}{
     It also used a built-in case-control design. From March 1, 2012, to May 31, 2019, \textbf{all breast cancer surgeries are routine}. After discharge, the surgeon has also evaluated three times \textbf{more than 30 days of the patient week}.
}

%-------------------------------------------
\subsection{Case 4: Citations to non-existent literature}

\onecase{Blockchain financial development based on FPGA and Convolutional Neural Network}{Not included in a volume, online since November 2020, \href{https://doi.org/10.1016/j.micpro.2020.103492}{doi:10.1016/j.micpro.2020.103492}.}

This article contains the following tortured phrases (Tab.~\ref{tab:tortured}):
\emph{profound neural organization} (in lieu of \emph{deep neural network}), 
\emph{fake neural organization} (in lieu of \emph{artificial neural network}), 
\emph{counterfeit neural organization} (in lieu of \emph{artificial neural network}),
\emph{human-made consciousness} (in lieu of \emph{artificial intelligence}).\\[6pt]

The reference list contains non-existent or unidentifiable items. The hyperlinks provided in the pdf (and reproduced below) are either broken or leading to unrelated publications:
\faketbox{Case 4 reference section -- GPT detector score 97.69\%}{
\begin{itemize}
    \item \hyperlink{http://refhub.elsevier.com/S0141-9331(20)30644-X/sbref0004}{[4]}: T.D. Chaudhry, Gauche, to decipher the forecast of the instability of the Indian financial exchange, utilizes a \textbf{counterfeit neural organization} with various information sources and yields, J. Estimation 120 (8) (2016) 7–15. \textbf{It applies to.}
    \item \hyperlink{http://refhub.elsevier.com/S0141-9331(20)30644-X/sbref0005}{[5]}: H. Moncada, M.H. Monhada, M. Esfandiari, Utilizing a \textbf{counterfeit neural organization}, J. Econ. Class Stock File Fig. 21 (41) (2016) 89–93. \textbf{Fund will be logical.}
    \item \hyperlink{http://refhub.elsevier.com/S0141-9331(20)30644-X/sbref0011}{[11]}: W. Melody, W. Tune, W. Song, R.F. expectation, \textbf{ACM bilayer neural organization system transformer}. The executives, Inf. Framework. 7 (4) (2017) 1–17.
    \item \hyperlink{http://refhub.elsevier.com/S0141-9331(20)30644-X/sbref0013}{[13]}: G. Kaur’s, J. Dahl, R.K. Ha, Used to foresee the mix with the BSE list in any event adjusted \textbf{OWA administrator ANFIS fluffy C-}, science, Neuropsychol. Rev. 122 (2016) 69–80.
\end{itemize}
}

Additionally, the related work section attributes ref. [4] to `Kiyoshi Erwang' whereas the references section gives `T.D. Chaudhry, Gauche.' We were unable to determine whether the identity of `Kiyoshi Erwang' is real or not.\\

The \emph{related work} section starts with the following text:
\faketbox{Case 4, \emph{related work} -- GPT detector score 98.60\%}{
The strategies for \textbf{human-made consciousness} has been invited by an ever-increasing number of residents. The \textbf{human-made consciousness} technique that speaks to exploring a \textbf{neural organization} has been created at an exceptional rate [1]. \textbf{Such a business expectation, Assessment of scores, business misfortune forecast, these fields, for example vision and control framework, has been generally utilized [2].}
}

%-------------------------------------------
\subsection{Case 5: Referring to a theorem that is never introduced}

\onecase{Ecological landscape planning and design based on the Internet of Things system and VR technology}{Not included in a volume, online since November 2020, \href{https://doi.org/10.1016/j.micpro.2020.103431}{doi:10.1016/j.micpro.\allowbreak{}2020.103431}.}

This article contains the following tortured phrases (Tab.~\ref{tab:tortured}):
\emph{organization association} (in lieu of \emph{network connection}), 
\emph{information distribution center} (in lieu of \emph{data warehouse}).

The introduction contains a reference to {\emph{Theorem 1.2}}, which does not exist within the paper. Most variables in mathematical formulas across the paper are not introduced in any way, their meaning remain unclear from the context. 

Figure 2 in the paper is identical to Figure 6 from \href{https://doi.org/10.1016/j.scib.2019.07.004}{doi:10.1016/j.scib.2019.07.004}. No acknowledgement for image reuse was provided.

The statement of author's research interests appears odd: \emph{``His research interests include Chinese calligraphy and fine arts, and the comparison of Chinese and Western arts.''} The title of the paper is \emph{``Ecological landscape planning and design based on the Internet of Things system and VR technology''}.

An excerpt from the abstract is provided below as an example of language irregularities in the paper.

\faketbox{Case 5, abstract -- GPT detector score: 99.96\%}{Intelligent, real-time, is \textbf{essential to low-cost}, the planning and design of a distributed ecosystem to understand, to manage the rapidly changing ecosystem. However, in the era of big data, most new technology, especially in remote areas of fragile ecosystems, \textbf{it is not introduced to the planning and design of the business ecosystem}. Innovative by using the Internet of things technology eco-system in the smart device planning and design and control system in the development and isolated environment, to establish the eco-system of the prototype, \textbf{internet of things (IoT) technology it is introduced}.}

%-------------------------------------------
\subsection{Case 6: Abstracts of other papers rewritten in a tortuous way}

\onecase{New technology application in logistics industry based on machine learning and embedded network}{Published in volume 80 of February 2021, \href{https://doi.org/10.1016/j.micpro.2020.103596}{doi:10.1016/j.micpro.2020.103596}.}

This article contains the following tortured phrases (Tab.~\ref{tab:tortured}):
\emph{organization association} (in lieu of \emph{network connection}), 
\emph{huge information} (in lieu of \emph{big data}),
\emph{arbitrary timberland} (in lieu of \emph{random forest}).

In the `Materials and method' section, one can read \emph{``FedEx and Uninterruptible Power Supply (UPS), has become two recipients and supporters of improving transport and coordination,''} the abbreviation UPS is obviously incorrect as it should be ``United Parcel Service.''

The related work section seems to be a concatenation of automatically re-written abstracts. This text could result from a back and forth automatic translation or the output of an unspecified re-writing tool:\\

\noindent
\begin{minipage}{0.48\textwidth}
\truetbox{Abstract of reference [13] in Case 6 references section -- GPT detector score: 0.02\%}{This work includes processing and classification of tweets which are written in Turkish language. \textbf{Four different sector} tweet datasets are vectorized with Word Embedding model and classified with Support Vector Machine and \textbf{Random Forests classifiers} and results have been compared. We have showed that sector based tweet classification is more successful \textbf{compared to general tweets}. Accuracy rates for Banking sector is 89.97\%, for Football 84.02\%, for Telecom 73.86\%, for Retail 63.68\% and for overall 74.60\% have been achieved.}
\end{minipage}
\hfill
\begin{minipage}{0.48\textwidth}
\faketbox{Questionable text in case 6, rewritten from [13] -- GPT detector score: 59.20\%}{This work contains the preparation and characterization of tweets written in Turkish. Tweet information sets a vector of \textbf{four different offices}, contrasted and the outcomes and the installed model and grouping support vector machine and the \textbf{arbitrary timberland arrangement} in Word. Area-based tweet arrangement of, \textbf{contrasted with the overall mumble}, has bpreparationstrated to be moderately effective. The exactness rate for the financial area 89.97 percent, soccer 84.02 percent, 73.86 percent for correspondence, it has been made 74.60 percent of the absolute of the 63.68 percent for retail.}
\end{minipage}

\noindent
\begin{minipage}{0.48\textwidth}
\truetbox{Abstract of reference [12] in case 6 references section -- GPT detector score: 0.22\%}{This study presents a comparison of different \textbf{deep learning methods} used for sentiment analysis in Twitter data. In this domain, deep learning (DL) techniques, which contribute at the same time to the solution of a wide range of problems, gained popularity among researchers. Particularly, two categories of neural networks are utilized, \textbf{convolutional neural networks (CNN)}, which are especially performant in the area of image processing and recurrent neural networks (RNN) which are applied with success in \textbf{natural language processing (NLP)} tasks. In this work we evaluate and compare ensembles and combinations of CNN and a category of RNN the long short-term memory (LSTM) networks.}
\end{minipage}
\hfill
\begin{minipage}{0.48\textwidth}
\faketbox{Case 6 text rewritten from its ref. [12] -- GPT detector score: 96.79\%}
{In this investigation, look at the different \textbf{profound learning strategies} for information notoriety examination of Twitter. Around there, simultaneously profound Deep learning (DL) innovation to add to the arrangement of a wide scope of issues has been invited by analysts. In particular, two classifications of neural organization uses typically picture preparing and \textbf{Neuro-Linguistic Programming (NLP)}, a neurstrategiesizatithe on \textbf{Cable News Network (CNN)} convolution is applied to the region of the repetitive neural organization Recurrent Neural Network (RNN) with the undertaking.}
\end{minipage}

\clearpage

%-------------------------------------------
\subsection{Case 7: Citing items missing from the reference list}

\onecase{Simulation of football based on PID controller and BP neural network}{Published in volume 81 of March 2021, \href{https://doi.org/10.1016/j.micpro.2020.103695}{doi:10.1016/j.micpro.2020.103695}}

While the reference list contains only 15 items, labelled [1] through [15], the section \emph{Related works} contains citations to items [16] and [17], as below:

\faketbox{Case 7, citations to items [16] and [17] missing from the reference list -- GPT detector score: 99.98\%}
{Calculation surveys that anticipate football results Competitions dependent on {\bf neural organizations}. Study the capacity to make forecasts dependent on broad variables chose to utilize the neural organization [16]. Because of test research on neural organization properties, the expectation precision accomplished. A productive and speedy approach {\bf to circulatory strain’s neural organization is utilized by football refs to help refs their actual quality and quality}. Partitioned into two classifications: Or can’t be against the mediator. This neural organization created under Boy Lab Windows, and its cordial U.I. created under Visual Basic. The Neural Network Method utilizes the past stage official point at the World Cup to foresee the two groups’ speed that won the football competition [17].}

We further provide an excerpt from the \emph{Introduction} to further highlight irregularities in grammar and vocabulary of the paper. It is remarkable that the GPT detector score of this fragment is very low. This kind of examples highlight the actual limitations of deep learning methods by questioning the lack of explanation regarding the computed results.

\faketbox{Case 7, \emph{Introduction} -- GPT detector score 1.50\%}
{As the \textbf{climate changes}, the ball’s situation comparative with the robot’s case is continually evolving. Subsequently, it is essential to apply a control framework for robots \textbf{to peruse these dynamic ecological conditions}. One of them is the utilization of \textbf{dark coherent regulators} in conduct based control and football robot route frameworks.}

%-------------------------------------------------------------------
\section{Potential sources of problematic papers}\label{sec:discussion}
This section discusses the  sources we suspect are involved in churning problematic papers out: paper mills and Spinbot-like software.

\subsection{Paper mills: Template-based massive production of papers}
We suspect papers mills \citep{ElseAndVanNoorden2021} to have produced part of the problematic papers we analysed. Several recurring features shared by most questionable papers from \emph{Microprocessors and Microsystems} may indicate that they come from a single source:
\begin{itemize}
    \item Similar composition of the papers consisting of five sections named (with slight variations) \emph{Introduction}, \emph{Related work}, \emph{Materials and methods}, and \emph{Results and discussion}, \emph{Conclusion}. This composition is not so common for papers published before volume~80, or even for papers in volumes~80--83 with longer duration of editorial assessment.\\[2pt]
    
    \item Most questionable papers that we inspected share the same typical set of colours used for diagrams: light blue, orange, grey, yellow, and blue. This suggests that the same software was used to prepare the papers. However, this feature is not common even for all questionable papers.\\[2pt]
    
    \item We share a subjective impression that there is little variability in the way the images and tables are prepared. For instance, the use of block diagrams is outstandingly common. We expect that presence of non-standard images in these papers will often indicate unacknowledged reuse.
\end{itemize}

In addition, let us note that the observed changes in the operational mode of the journal, most notably, the increased output, bear some resemblance to how \emph{hijacked journals} operate \citep{JalalianAndDadkhan2015,Abalkina2021}.

%-------------------------------------------------------------------
\subsection{Spinbot: Article Spinning, Text Rewriting, Content Creation Tool}
    Searching the web for ‘text reformulation’ we stumbled upon \href{http://spinbot.com}{spinbot.com}, introduced as ``a free, automatic article spinner that will rewrite human readable text into additional, intelligent, readable text.''
    The Wayback Machine of the Internet Archive has records of this website offering this service for a decade now.\footnote{See the archive of January 28, 2011 at \url{https://web.archive.org/web/20110128/http://spinbot.com/}. Note that the feature called `Spin Any Language to Any Language' once present is not available anymore.}
    The \href{https://web.archive.org/web/20210617101807/https://spinbot.com}{web page} offers to ‘rewrite’ a text of up to 10,000 characters for free.
    No information is provided regarding the technology Spinbot uses, no computer code is available.
    Two \href{https://web.archive.org/web/20210617101807/https://spinbot.com/Pricing}{paid options} are proposed.
    First, a paid subscription allows customers to use Spinbot without ads or captchas for \$10 a month, \$50 for 6 months, or \$75 a year.
    Second, a developer can buy ‘Spin Credits’ that correspond to a number of API calls, prices ranging between \$5 for 1,000 credits and \$2,000 for 500,000 credits.
    Multiple websites such as \href{https://paraphrasing-tool.com}{paraphrasing-tool.com} and \href{https://free-article-spinner.com}{free-article-spinner.com} claim to be ‘powered by the Spinbot API.’
    There is no personal information available about the designer of Spinbot yet the \href{https://blog.spinbot.com}{Spinbot Blog} advertises the \href{https://mrgreenmarketing.com}{Mr. Green Marketing, LLC} company based in Kansas City, USA.
    
    Studies of academic writing use the term `spin' to describe authors' attempt to present a more positive description of a technique or drug in health sciences than the data actually support \citep{BoutronEtAl2014,BoutronAndRavaud2018}.
    What Spinbot performs, however, is less complex: it replaces words with synonyms.
    For instance, the text `big data' is transformed into `enormous data' or `huge data' or `large data' when running Spinbot multiple times.
    The term `artificial intelligence' is spun as `counterfeit consciousness' or `man-made brainpower' or `computerized reasoning.'
    Feeding the phrases in the ‘Correct wording expected’ column of Tab.~\ref{tab:tortured} to Spinbot we were able to reproduce the associated `tortured phrases.' Different executions of Spinbot with ``The road to hell is paved with good intentions.'' as input yielded:
    \begin{itemize}
        \item The road to hell is paved with good intentions.
        \item The way to damnation is cleared with sincere goals.
        \item The way to hellfire is cleared with honest goals.
        \item The way to hellfire is cleared with well meaning goals.
        \item The way to damnation is cleared with well meaning goals.
    \end{itemize}
    
%-------------------------------------------------------------------
\section{Conclusion and Call for Action}\label{sec:conclusion}
We discovered a number of tortured phrases in the scientific literature, mainly in Computer Science.

We further studied one specific journal, \emph{Microprocessors and Microsystems}, affected by the phenomenon. Our study revealed significant, likely questionable, changes in the journal's operational mode. These changes did not attract much attention, despite being given out by various hints, including:
\begin{itemize}
    \item abrupt drop of the average / median duration of editorial assessment;
    \item abrupt surge in the number of articles accepted;
    \item acceptance of evidently synthetic texts; 
    \item unexpected author affiliations and/or research interests (in authors' biographies) and/or research background outside of the scope of the venue (e.g., `school of musicology' in a venue on microprocessors).
\end{itemize}

We specifically discussed the issues with 7 cases covering 8 papers that we are also reporting on PubPeer \citep{BarbourAndStell2020}.
As of 17 June 2021, none of the 1,078 papers had been commented on PubPeer (\supplmat), suggesting that the issues we found went unnoticed.

No systematic screening of the papers containing tortured phrases has been performed to date.
Nevertheless, we estimate \emph{Microprocessors and Microsystems} accepted around 500 questionable articles: 389 papers with short duration of editorial assessment in volumes 80--83, plus additional papers not yet included in a volume.
As of June 25, 2021 there were 225 such articles queued `in press' which are `accepted, peer reviewed articles that are not yet assigned to volumes/issues, but are citable using DOI.'

Our study revealed that multiple other venues also published papers with tortured phrases (Fig.~\ref{fig:microproDimensions}) and abstracts with high GPT detector scores (Tab.~\ref{tab:gptEls2021}).
Tailoring the fingerprint--query approach used in \citep{CabanacAndLabbe2021} is a promising way to comb the literature for tortured phrases.\footnote{See the screening and assessment results for grammar `tortured' at \url{https://www.irit.fr/~Guillaume.Cabanac/problematic-paper-screener}}
Preliminary probes show that several thousands of papers with tortured phrases are indexed in major databases.
While we managed to identify and retro-engineer several tortured phrases in Computer Science, other tortured phrases related to the concepts of other scientific fields are yet to be exposed.\\[12pt]

\noindent All in all this work is a call to action and we therefore:
\begin{itemize}
    \item Encourage other members of the scientific community to extend our findings. We also welcome any effort towards deeper case-by-case analysis of papers in various venues.\\[-8pt]
    
    \item Expect that the relevant parties (Elsevier, COPE, Clarivate Analytics, etc.) initiate an impartial, efficient, transparent and wide investigation, should our concerns be grounded.  We believe that the irregularities we raised in this study may fall within the scope of the COPE's guideline/flowchart \emph{Systematic manipulation of the publication process} \citep{COPE2018}. \\[-8pt]
    
    \item Suggest that researchers and, especially, publishers monitor the publishing ecosystem for various hints indicating unusual publication activities (see below). However, we emphasise that hints alone do not mean that misconduct is happening, therefore an analysis of individual papers should be performed in order to support or refute any concerns.
\end{itemize}

We wish to broaden the discussion about whether software and natural language models are welcome to generate or modify scientific texts. It is of tremendous importance to provide the detection method that goes along. Such detection methods should be well characterised with regards to both false positives (type~I error) and false negatives (type~II error). The detection method should also provide a rationale for its decision in line with the current expectations for `explainable artificial intelligence.'

Attempts to automatically detect synthetic texts will benefit from open abstracts\footnote{\url{https://i4oa.org}} \citep{Schiermeier2020} and their indexing by various academic search engines such as Dimensions \citep{HerzogEtAl2020}.
Screening pipelines \citep[e.g., ][]{WeissgerberEtAl2021a} may include such detection software and run it ahead of peer review. While probably useful in short-to-medium term perspective, we fear that any screening initiative is likely to provoke an arms race.

We note, however, that peer review --- or rather initial editorial screening --- should have detected and filtered out the most blatant examples of synthetic texts; its failure to do so should be analysed.

In our strong opinion, the root of the problems discussed in this work is the notorious \emph{publish or perish} atmosphere \citep{Garfield1996} affecting both authors and publishers.
This leads to blind counting and fuels production of uninteresting (and even nonsensical) publications.\\[12pt]

\noindent\textbf{Update as of July 12, 2021:} \emph{Retraction Watch} reports Elsevier issued Expressions of Concern for six special issues of \emph{Microprocessors and Microsystems} \citep{Marcus2021}.  It is not clear whether regular papers (see Tab.~\ref{tab:lessThan30d}) will be assessed.

%-------------------------------------------------------------------
\section*{Appendix: List of supplementary materials}\label{apx:supplmat}
We release supplementary materials on Zenodo under \href{https://doi.org/10.5281/zenodo.5031935}{doi:10.5281/zenodo.5031935} for transparency and reproducibility concerns.

% Actions to be taken by the publisher as per their own policy:
%\url{https://www.elsevier.com/about/policies/article-withdrawal}
%\url{https://www.elsevier.com/editors/perk/allegations-of-research-errors-and-fraud} 

\begin{acknowledgement}
We thank \href{https://pubpeer.com}{PubPeer} for providing a forum to the  scientific community.
We thank Digital Science for making \href{https://dimensions.ai}{Dimensions} data available for scientometric research. A.M. would like to thank Maxim Panov (Skoltech) for help with initial reconnaissance of the subject. A.M. also acknowledges Mike Downes (independent researcher, Australia) who has already observed the phenomenon of recurring editorial timelines in predatory venues, although his post on the subject at \url{https://scholarlyoutlaws.com/} appears to be no longer accessible.
We are grateful to the colleagues who provided constructive comments and feedback on a previous version of this preprint: Frédérique Bordignon, Ophélie Fraisier-Vannier, Willem Halffman, Vincent Larivière, and François Portet.

\end{acknowledgement} 
 
%===============================================================================
\bibliographystyle{apacite}
\interlinepenalty=10000 % http://tex.stackexchange.com/a/51259
%\bibliography{generatedTexts}

\end{document}